\documentclass[%
 reprint,
 amsmath,amssymb,
 aps,
]{revtex4-2}

\pdfpageattr {/Group << /S /Transparency /I true /CS /DeviceRGB >>}
\usepackage{graphicx}
\usepackage{float}
\usepackage{soul}
\usepackage[normalem]{ulem}
\usepackage{amsmath,amssymb,amstext,dsfont,tikz,physics,mathtools,simpler-wick,stmaryrd}
\usepackage[colorlinks=true, allcolors=purple]{hyperref}

\usepackage{comment}
\usepackage{graphicx}
\usepackage{dcolumn}
\usepackage{bm}
\usepackage{bbm}
\usepackage{xcolor}
\usepackage{braket}

\newcommand{\oml}{\omega_{\mathrm{l}}}

\newcommand{\Bt}{B_{\mathrm{t}}}
\newcommand{\Ds}{\Delta_{\mathrm{s}}}
\newcommand{\Dc}{\Delta_{\mathrm{c}}}

\begin{document}

\preprint{APS/123-QED}

\title{Sideband Spectroscopy in the Strong Driving Regime:\\
        Volcano Transparency and Sideband Anomaly}

\author{Luka Antonić}
 \affiliation{Physics Department, Technion, Haifa
32000, Israel}
\email{lukaantonicx@gmail.com}
\author{Sergey Hazanov}%
\affiliation{Andrew and Erna Viterbi Department of Electrical Engineering, Technion, Haifa
32000, Israel}
\author{Sergei Masis}
\affiliation{Andrew and Erna Viterbi Department of Electrical Engineering, Technion, Haifa
32000, Israel}
\author{Daniel Podolsky}
 \affiliation{Physics Department, Technion, Haifa
32000, Israel}
\author{Eyal Buks}
\affiliation{Andrew and Erna Viterbi Department of Electrical Engineering, Technion, Haifa
32000, Israel}


\date{\today}

\begin{abstract}
We study the response of a spin to two crossed magnetic fields: a strong and fast transverse field, and a weak and slow longitudinal field. We characterize the sideband response at the sum and the difference of driving frequencies over a broad range of parameters. In the strong transverse driving regime, the emission spectrum has a characteristic volcano lineshape with a narrow central transparency region surrounded by asymmetric peaks. Next, we couple the spin to a nonlinear cavity that both drives and measures it. In a sufficiently slow longitudinal field, the emission spectrum exhibits anomalous behavior, where the resonances in both the right and left sidebands lie on the same side of the central resonance. The theoretical results are compared to the experimental measurement of the emission of substitutional nitrogen P1 and nitrogen-vacancy NV$^-$ defects in diamond.
\end{abstract}

\maketitle

\section{\label{sec:level1}Introduction}

In typical magnetic resonance experiments, the system is driven and measured at the same frequency, making it difficult to detect weak response signals amid the strong driving.  Another challenge appears under strong driving, where power broadening renders the spectrum featureless. 
In contrast, the sideband method probes the system at a frequency different from that used for driving, which enables implementing spectral filtering in the readout to eliminate the interference of the driving tone.  We consider a setup that combines a strong and fast transverse drive at angular frequency $\omega_{\mathrm{t}}$ with a weak and slow longitudinal drive at $\omega_{\mathrm{l}}$ (throughout the paper, the letter $\omega$ is used to denote angular frequencies). We will show that when $\omega_{\mathrm{l}}$ is larger than the decoherence rate $\gamma_2=1/T_2$, power broadening does not wash out the sideband spectra at $\omega_{\mathrm{t}}\pm\omega_{\mathrm{l}}$, but produces distinct features that we characterize.

In a seminal paper by Redfield~\cite{redfield1955}, the response to crossed fast transverse and weak longitudinal fields was studied in spin systems~\cite{redfield1955}. Since the oscillating longitudinal field leads to the reduction of magnetic polarization, the effect was named \emph{rotary saturation}. It was extended to optical systems~\cite{prior1977, prior1978}, and has potential applications in magnetic resonance imaging in medicine~\cite{witzel2008, ueda2018}. The longitudinal field is known to affect the Rabi frequency~\cite{saiko2010, saiko2012, saiko2014, glenn2013, yan2017, forster2015, greenberg2005, greenberg2007}, which is the rate of population oscillations in a transversely driven two-level system.

\begin{figure}
    \centering
    \includegraphics[width=.5\textwidth]{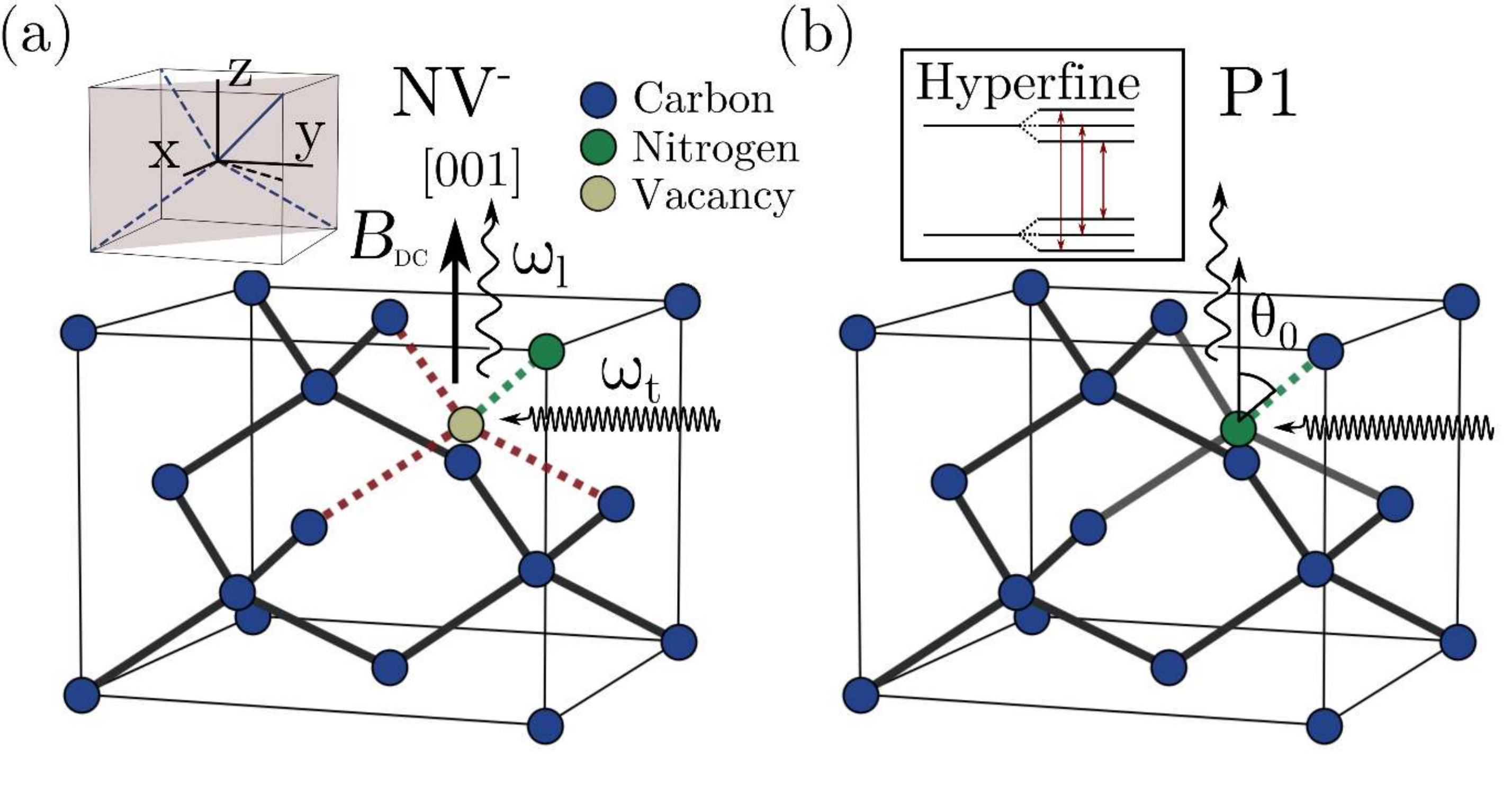}
    \caption{NV center (a) and P$1$ center in diamond (b). Both centers are subjected to DC magnetic field in the $[001]$ direction, drawn as the z-axis in the figures. This field sets the quantization axis. The slow radio frequency (RF) field ($\omega_{\mathrm{l}}$) is applied parallel to this axis, while the fast microwave field ($\omega_{\mathrm{t}}$) is perpendicular. Both centers have $C_{3v}$-symmetry.}
    \label{fig:diamond_balls}
\end{figure}

In this work, we present an experimental and theoretical study of the response of spin systems to crossed fields. We perform experimental measurements on two types of diamond defect: the negatively charged nitrogen-vacancy NV$^{-}$ center and the substitutional nitrogen N$_{\mathrm{S}}$ defect (also known as the P$1$ center or C-center). The NV center consists of the substitutional nitrogen and a neighboring vacancy in place of two carbon atoms, as shown in Fig.~\ref{fig:diamond_balls}(a). The center has been extensively studied both theoretically and experimentally  ~\cite{maze2011, doherty2013, balasubramanian2009}, and it is significant for emerging quantum technologies~\cite{togan2010}. By contrast, the P$1$ center, shown in Fig.~\ref{fig:diamond_balls}(b), appears when one of the carbons is exchanged with the nitrogen. This defect has a spin-$1/2$ and it was initially detected in electron-spin resonance experiments~\cite{smith1959}. 

In the experiment, a static magnetic field $B_{\mathrm{DC}}$ induces Zeeman splitting of the spins. 
On top of this field, we apply a weak, low-frequency longitudinal field at  $\omega_{\mathrm{l}}$ parallel to the dc field, as well as a strong, high-frequency transverse field with angular frequency $\omega_{\mathrm{t}}$ perpendicular to it. Under this two-tone driving, the spin oscillates at the sum and difference of the frequencies, acting as a frequency mixer. The oscillating spin emits radiation at central frequency $\omega_{\mathrm{t}}$ and at sideband frequencies $\omega_{\mathrm{t}}\pm\omega_{\mathrm{l}}$, which we then measure as the emission spectrum. We show that the sideband spectrum exhibits a more complex response than the central resonance, resolving system details more clearly, with promising applications in quantum sensing.

We complement our experimental measurements with a theoretical analysis of the sideband response. Specifically, we solve the dynamical equations of motion non-perturbatively in the strong transverse field, while treating the weak longitudinal field within linear response. This approach enables us to characterize the sideband spectrum across a range of regimes as the longitudinal frequency $\omega_{\mathrm{l}}$ and transverse amplitude $B_{\mathrm{t}}$ are varied. When $\omega_{\mathrm{l}}$ is smaller than the decoherence rate $1/T_2$, the spectrum is power-broadened and featureless.  As $\omega_{\mathrm{l}}$ exceeds $1/T_2$, distinct spectral regimes emerge, exhibiting rich structure absent in the power-broadened central resonance under strong driving. At intermediate transverse field amplitudes, the sideband reveals Rabi resonances that occur when the longitudinal field matches the Rabi frequency of the oscillating spin~\cite{bishop2009, li2013, rohr2014}. At very high amplitudes, a new spectral structure emerges: an asymmetric volcano lineshape with a narrow central dip, which we term the \emph{volcano transparency}.

In our analysis, we demonstrate that, for a broad class of linear theories, the right and left sidebands should be identical, apart from an overall inversion in the $\omega_\mathrm{t}$–$B_{\mathrm{DC}}$ plane. For instance, if the right sideband spectrum is blue-shifted relative to the frequency of the cavity used to drive the system, then the left sideband spectrum must necessarily be red-shifted.  This, however, is not observed experimentally. Instead, both sidebands are simultaneously blue-shifted, a phenomenon we refer to as the {\em sideband anomaly}. This anomaly serves as a sensitive indicator of nonlinearity in the system dynamics~\cite{levi2020}. To illustrate this theoretically, we demonstrate that the sideband anomaly can occur in systems with Kerr nonlinearity.  However, we note that Kerr nonlinearity is relatively weak in our system, and the source of the observed sideband anomaly remains unidentified.

Our paper is organized as follows.  We first describe the experiment in Sec.~\ref{sec:experiment} and introduce its theoretical description in Sec.~\ref{sec:setup}. In Section~\ref{sec:drivenSpin} we study the spin response at the dual driving by the strong and fast transverse magnetic field and weak and slow longitudinal field. In Section~\ref{sec:spin_cavity}, we show that sideband symmetry is guaranteed in the linear model. In Section~\ref{sec:nonlinear_cavity} we extend the model by adding a cavity with Kerr nonlinearity, demonstrating how it shifts the spectrum, where the shift strongly depends on the ratio of the longitudinal frequency and the cavity decay rate. We find that, if the longitudinal driving is slow, the sidebands get more blue-shifted than the central spectral line, and the right and left sidebands have different shifts if the cavity nonlinearity is strongly excited.

\section{The experiment}\label{sec:experiment}

\subsection{Experimental setup}
\label{sec:exp_setup} 

\begin{figure}
\begin{center}
\includegraphics[width=.49\textwidth, keepaspectratio]{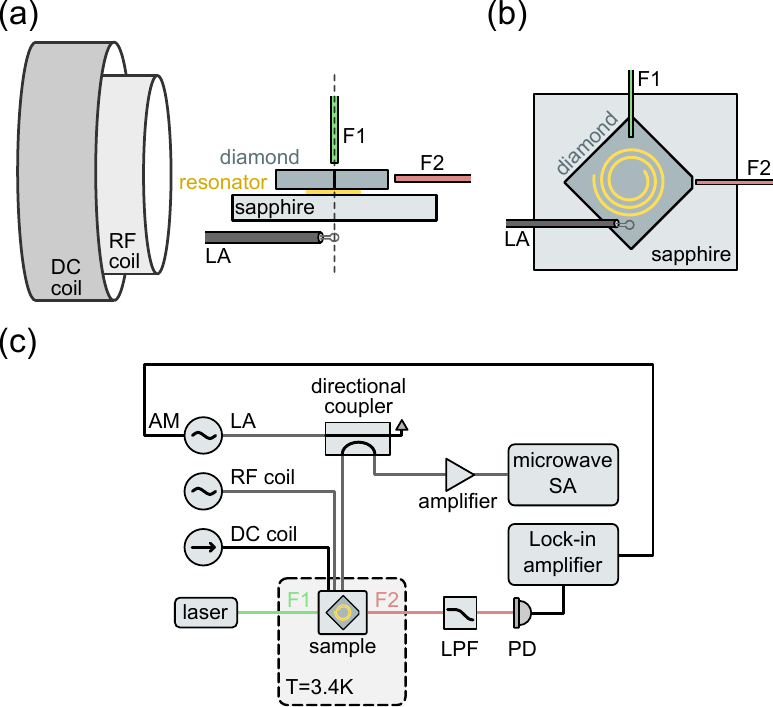}
\end{center}
\caption{(a) A sketch of the experimental setup with solenoids. (b) Top-down view of the central part of the setup. (c) Electrical and optical diagram with components described in Section~\ref{sec:exp_setup}. Black and gray lines label electrical connections, and red and green lines label optical connections. 
}%
\label{fig:FigSetup}%
\end{figure}

Figure~\ref{fig:FigSetup} shows the experimental layout inside the cryostat at a temperature of about $3.4\operatorname{K}$. A $\left[  110\right]$ diamond wafer, with NV-defect concentration of $1.23\times 10^{17} \mathrm{cm}^{-3}$, is glued to a
sapphire wafer. This wafer supports a niobium superconducting spiral resonator having frequency of $\omega_{\mathrm{c}}/\left(2\pi\right) =3.772\operatorname{GHz}$~\cite{Maleeva_064910, alfasi2018}, and cavity width $\gamma_{\mathrm{c}}\approx 8.5$MHz determined by fitting data to the Lorentzian function. A superconducting DC solenoid (see inset of Fig. \ref{fig:FigSetup} a) generates the static magnetic field, while an RF solenoid, with its axis parallel to the DC coil, provides the longitudinal drive at an angular frequency which we denote $\omega_{\mathrm{l}}$. The diamond crystal direction $\left[ 001\right]$ is aligned to be nearly parallel to the static magnetic field. A loop antenna (LA) that is placed beneath the spiral resonator
is employed for applying transverse driving in the microwave (MW) band, whose frequency we denote as $\omega_{\mathrm{t}}$. The role of the spiral resonator is twofold; it serves as a bandpass filter that passes frequencies close to the fundamental mode, but it also amplifies the signal from the loop antenna and from the defects.

The static field $B_{\mathrm{DC}}$ affects the Zeeman splitting of the defects.  Although both centers have four possible orientations, applying the field along a $\langle100\rangle$ direction, which corresponds to angle $\theta_0=\mathrm{arccos}\left(1/\sqrt{3}\right)$ relative to $\langle111\rangle$ bonds, causes identical response for all four orientations, see Fig.~\ref{fig:diamond_balls}.

Two multi-mode optical fibers, F1 and F2 in Fig. \ref{fig:FigSetup}(a), are coupled to the diamond sample. Laser light at $532\operatorname{nm}$ (which is delivered to the sample by fiber F1) gives rise to optically induced spin polarization \cite{Robledo_025013, Redman_3420}. The experimental setup allows implementing both methods of optical detection of magnetic resonance (ODMR) and cavity-based detection of magnetic resonance (CDMR). The photoluminescence (PL) emitted from the diamond, which is delivered by the multi-mode fiber F2, is probed using a long pass filter (LPF) and a photo detector (PD). The PD signal is measured using a lock-in amplifier (LIA). The MW signal generator, which is employed for applying transverse driving with the LA, is amplitude modulated (AM) using a low-frequency signal generated by the LIA.

The scheme of the sideband spectroscopy setup is shown in the upper part of Fig. \ref{fig:FigSetup}(b). A directional coupler (DC) is employed for delivering incoming and outgoing LA signals. The incoming signal, which has the frequency $\omega_{\mathrm{t}}$, is generated by an MW function generator. An amplifier and a spectrum analyzer (SA) are used to probe the signal outgoing from the LA. An RF signal generator having frequency $\omega_{\mathrm{l}}$ is employed for applying longitudinal driving to the RF solenoid. The defect's nonlinear response mixes transverse and longitudinal driving tones, producing the sideband peaks at frequencies $\omega_{\mathrm{t}}\pm n\omega_{\mathrm{l}}$, where $n$ is an integer. We record the sideband amplitudes with SA and, throughout the paper, we focus only on the first right and left sidebands with $n=\pm 1$.

\subsection{Measurements on P1 centers}

\begin{figure*}
    \centering
\includegraphics[width=.95\textwidth]{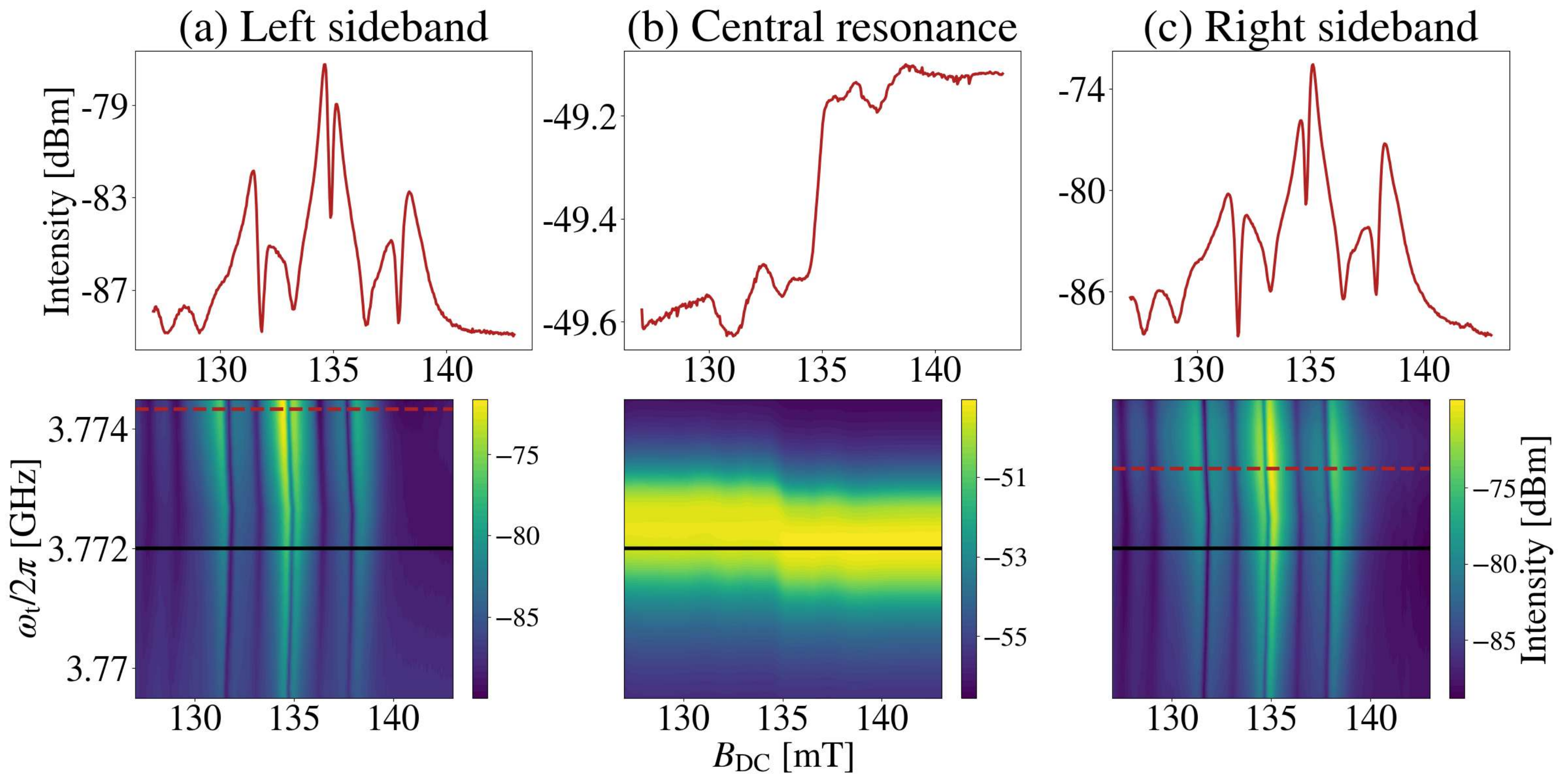}
    \caption{P1 sidebands, spectrum analyzer power measurement; intensity is measured in dBm units. (a) Left sideband, $\omega_{\mathrm{t}}-\omega_{\mathrm{l}}$,  (b) central peak at $\omega_{\mathrm{t}}$, and (c) right sideband, $\omega_{\mathrm{t}}+\omega_{\mathrm{l}}$. The horizontal axis
corresponds to the static magnetic field magnitude $B_{\mathrm{DC}}$, and the vertical axis corresponds to
the microwave driving frequency $\omega_{\mathrm{t}}/(2\pi)$. The $\mathrm{RF}$ drive
frequency is set to $1\,\mathrm{MHz}$. The overlaid black lines in the lower figures mark the cavity frequency, which is also the frequency at which the upper lineshapes are extracted. Red dashed lines mark the approximate positions of the sideband resonances. The three volcanoes correspond to three different hyperfine transitions.}
    \label{fig:experimental_volcano}
\end{figure*}

Figure \ref{fig:experimental_volcano} presents the measurements on the P1 center. The transverse field is amplified by the spiral resonator at frequency $\omega_{\mathrm{c}}/\left(
2\pi\right)=3.772\operatorname{GHz}$ (discussed in more detail in Sec.~\ref{sec:exp_setup}). The resonator also amplifies the radiation emitted from the spin, enhancing the sideband signal.
We resolve the hyperfine splitting induced by the interaction of the electron spin with the spin-1 nitrogen nucleus.  This results in three volcano-shaped transparencies in the top row, which correspond to the three hyperfine transitions displayed schematically in Fig.~\ref{fig:diamond_balls}(b) (see Appendix \ref{app:p1nv} for more details).
While the defect's signature is weak in the central response, it emerges as a well-defined lineshape in the sideband spectrum. The peaks surrounding the central transparency are asymmetric (one is higher than the other), and this asymmetry is reversed between the left and right sidebands. In Sec.~\ref{sec:drivenSpin}, we show theoretically how this spectral line emerges through the competition between Rabi resonances and power broadening.

The bottom row of Fig.~\ref{fig:experimental_volcano} displays measurements performed at various transverse frequencies $\omega_{\mathrm{t}}$. The black line marks the cavity frequency.  One expects the positions of the resonances in the right and left sidebands to be shifted in opposite directions relative to the cavity frequency.  Instead, both sideband signals are strongest at a blue shift from the cavity frequency. We call this phenomenon the \textit{sideband anomaly}.

\subsection{Measurements on NV centers}

\begin{figure*}
    \centering
    \includegraphics[width=.95\textwidth]{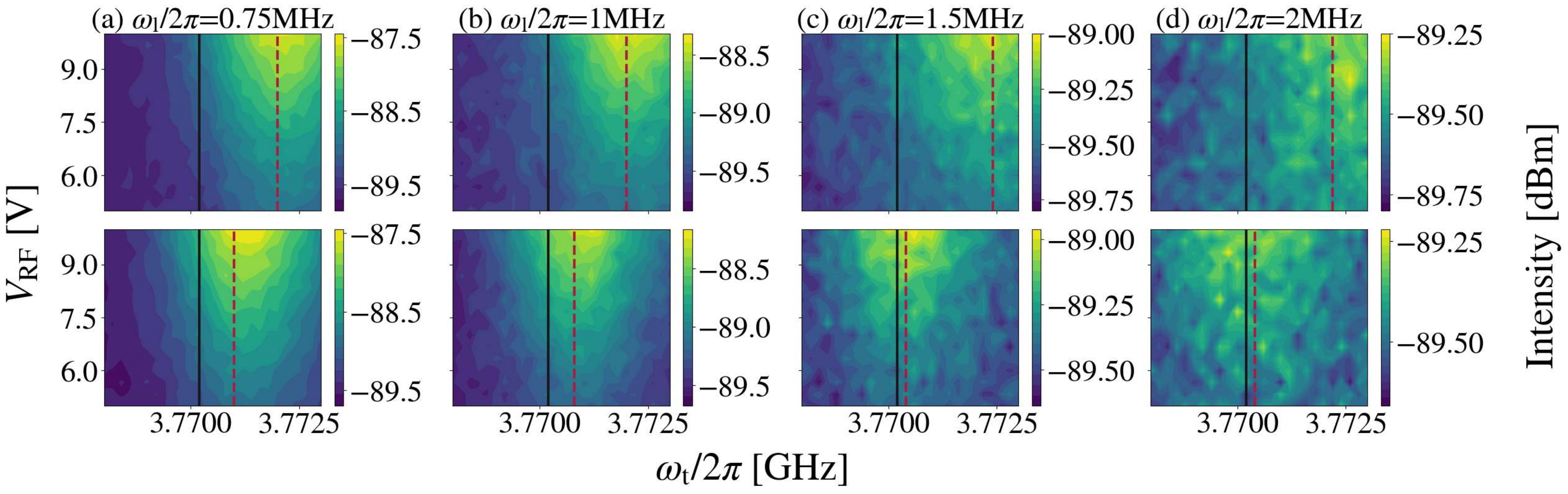}
    \caption{The sideband measurements of the NV center as a function of the longitudinal drive strength $B_{\mathrm{l}}$ (corresponding to the voltage on the RF coil $V_{\mathrm{RF}}$) and the transverse driving frequency $\omega_{\mathrm{t}}$ for different longitudinal driving frequencies $\omega_{\mathrm{l}}$. The top row represents the left sideband data (at the frequency $\omega_{\mathrm{t}}-\omega_{\mathrm{l}}$), and the bottom row are the right sideband data (at the frequency $\omega_{\mathrm{t}}+\omega_{\mathrm{l}}$). The black vertical line is the cavity frequency $\omega_{\mathrm{c}}$ as obtained from the experimental measurements. The red dashed lines show the frequency $\omega_t$ where the maximal signal, averaged over all $B_{\mathrm{l}}$ values, is observed. The sideband anomaly occurs for all of the longitudinal frequencies, but the distance between the sideband resonances increases with the increasing $\omega_{\mathrm{l}}$ (which also occurs with the Kerr nonlinearity, see Sec.~\ref{sec:nonlinear_cavity}).}
    \label{fig:nv_rf_frequency} 
\end{figure*}

In Figure 4, we show measurements on NV centers. The horizontal axis is the transverse driving frequency, $\omega_{\mathrm{t}}$, and the vertical axis is the RF coil voltage, which controls the strength of the longitudinal drive $B_{\mathrm{l}}$. The top row displays the left-sideband signal, while the bottom row shows the right sideband. Each column contains both sidebands for a specific longitudinal frequency, $\omega_{\mathrm{l}}$. The signals are more pronounced with increasing longitudinal driving.  Such behavior is expected when the driving is weak and justifies our later analysis, where we treat the longitudinal drive in linear response.  By contrast, for strong longitudinal drive, the sideband signal is expected to become non-monotonic and follow a more complicated dependence involving Bessel functions~\cite{shevchenko2018}.

The NV data confirm the sideband anomaly seen in P1‑center measurements. A vertical black line marks the cavity frequency, and in all cases, both sidebands lie to its right. In Section~\ref{sec:spin_cavity}, we demonstrate that a purely linear system would place the left and right sidebands on opposite sides of the cavity frequency (indicated by the black lines in the figure).

Furthermore, as $\omega_{\mathrm{l}}$ increases, the left and right sidebands show asymmetric behavior.  As shown in Sec.~\ref{sec:nonlinear_cavity}, Kerr nonlinearity gives rise to similar behavior.  This illustrates that nonlinearity can indeed cause the sideband anomaly. However, in the current experiment, Kerr nonlinearity is far too small to explain the observed anomaly, and its origin remains unexplained.

\section{\label{sec:setup}Theoretical Model}

We now introduce the theoretical spin-cavity model used to describe the experimental setup of the preceding section. The spin is placed in a constant magnetic field ${\bf B}_\mathrm{DC}=B_\mathrm{DC} \hat{z}$, leading to a Zeeman splitting $\omega_\mathrm{s}$.  The spin is driven by a weak and slowly oscillating longitudinal (parallel to $\hat{z}$)  field.  At the same time, the cavity is driven by a strong and fast oscillating field.  Due to the spin-cavity interaction, the fast cavity oscillations are transferred to the spin, which therefore feels the effects of both drives.
Then, as a result of frequency mixing, the spin emits radiation at the sum and difference of the two driving frequencies.  The emitted radiation is amplified by the cavity.  The resulting spectrum, which is a function of two frequencies, reveals a rich structure that we wish to explore. 

We introduce the Hamiltonian:
\begin{equation}
    H = H_\mathrm{s} + H_\mathrm{c} + H_\mathrm{int}\ .
\end{equation}
The spin part of the Hamiltonian is
\begin{equation}
    H_\mathrm{s} = \frac{\omega_\mathrm{s}+B_\mathrm{l}\cos{\left(\omega_\mathrm{l} t\right)}}{2}\sigma_\mathrm{z}\, ,
\end{equation}
where $\sigma_{\mathrm{z}}$ is Pauli matrix. The cavity is driven at frequency $\omega_\mathrm{t}$ and it has a Kerr nonlinearity,
\begin{equation}
    H_\mathrm{c} = \omega_\mathrm{c}a^{\dagger}a+\frac{K_\mathrm{c}}{4} a^{\dagger}a^{\dagger}aa+\frac{g_\mathrm{P}}{2}\left(a^{\dagger}e^{-i\omega_\mathrm{t} t}+ae^{i\omega_\mathrm{t} t}\right)\, ,
\end{equation}
where $a$ is the cavity annihilation operator. The spin couples to the cavity through a Jaynes-Cummings interaction,
\begin{equation}
    H_{\mathrm{int}} = \Omega \left(a^{\dagger}\sigma_-+a\sigma_++a^{\dagger}\sigma_++a\sigma_-\right)\ .
\end{equation}
This interaction provides the feedback mechanism between the spin and cavity as it transfers energy between them. 
Due to the form of the interaction, the cavity drive at frequency $\omega_\mathrm{t}$ is transferred to the spins as a field that is transverse to $\hat{z}$.

We can simplify the Hamiltonian by working in the rotating frame. The unitary operator that transforms the system into this frame is
\begin{equation}
    P(t) = e^{i \omega_{\mathrm{t}}\left(a^{\dagger}a-\frac{1}{2}\left(1-\sigma_\mathrm{z}\right)\right)t}\ .
\end{equation}
The Hamiltonian that evolves states and operators in the new frame is
\begin{equation}
    H'(t) = P(t)H(t)P^{\dagger}(t)-iP(t)(\partial_t P^{\dagger}(t))\ ,
\end{equation}
while the time-dependent observable $A'(t)$ in the rotating frame is related to the observable $A$ in the original frame through
\begin{equation}
    A'(t) = P(t)AP^{\dagger}(t)\ .
\end{equation}
The third and fourth terms in the expression for $H_{\text{int}}$ are rotating fast in the direction opposite to the frame, and they will be neglected in the equations of motion. The Hamiltonian is then, up to constant terms,
\begin{equation}
\begin{split}
    &H' = \Delta_{\mathrm{c}} a^{\dagger} a + K_\mathrm{c} a^{\dagger} a^{\dagger} a a+\frac{g_\mathrm{P}}{2}\left(a+a^{\dagger}\right)\\
    &+\frac{\Delta_\mathrm{s}+B_{\mathrm{l}}\cos{\left(\omega_{\mathrm{l}}t\right)}}{2}\sigma_\mathrm{z}+\Omega(a^{\dagger}\sigma_-+a\sigma_+).
\end{split}
\end{equation}
Here, we introduced two parameters,
\begin{eqnarray}
    &\Delta_\mathrm{c}=\omega_{\mathrm{c}}-\omega_{\mathrm{t}}, \\
    &\Delta_\mathrm{s}=\omega_{\mathrm{s}}-\omega_{\mathrm{t}},
\end{eqnarray}
which quantify the detuning of the transverse drive from the cavity and the spin, respectively. We study the emission spectrum as function of these two parameters. 


In addition to unitary evolution, the system experiences dissipation, which we model by the following relaxation parameters: $\gamma_\mathrm{c}$ for the cavity, $\gamma_2$ for the spin decoherence rate, and $\gamma_1$ for the spin dissipation rate, which tends to relax it to the equilibrium spin polarization $\sigma_\mathrm{z}^{\mathrm{eq}}$. In experiments that use optical spin polarization, the effective dissipation parameters can be very different from the intrinsic dissipation parameters of the system ~\cite{alfasi2018}. 

We write the equations of motion for the expectation values of $\sigma_{\mathrm{z}}$, $\sigma_+$, and $a^\dagger$:
\begin{subequations}
\begin{equation}
\label{eq:original_systema}
i\frac{d\langle\sigma_\mathrm{z}\rangle}{dt}= 2\Omega\left(\langle a\rangle\langle\sigma_{+}\rangle-\langle a^{\dagger}\rangle\langle\sigma_{-}\rangle \right)-i\gamma_{1}\left(\langle\sigma_{\mathrm{z}}\rangle-\sigma_{\mathrm{z}}^{\mathrm{eq}}\right),
\end{equation}
\begin{equation}\label{eq:original_systemb}
i\frac{d\langle\sigma_+\rangle}{dt} = -(\Delta_{\mathrm{s}}+B_\mathrm{l}\cos{\left(\omega_\mathrm{l} t\right)}+i\gamma_2) \langle\sigma_{+}\rangle+\Omega \langle a^{\dagger}\rangle\langle\sigma_\mathrm{z}\rangle,
\end{equation}
\begin{equation}\label{eq:original_systemc}
i\frac{d\langle a^{\dagger}\rangle}{dt} = -\left(\Delta_{\mathrm{c}}+i\gamma_\mathrm{c}\right)\langle a^{\dagger}\rangle-K_\mathrm{c}|\langle a^{\dagger}\rangle|^2\langle a^{\dagger}\rangle
-N_\mathrm{S}\Omega\langle\sigma_+\rangle -\frac{g_\mathrm{P}}{2}.
\end{equation}
\end{subequations}
The corresponding equations for $\sigma_-$ and $a$ can be obtained from the above by complex conjugation, by using $\langle\sigma_{\mathrm{z}}\rangle=\langle\sigma_{\mathrm{z}}\rangle^*$, $\langle \sigma_-\rangle=\langle \sigma_+\rangle^*$ and $\langle a\rangle=\langle a^\dagger\rangle^*$. In these equations, we used the mean-field approximation $\langle AB\rangle=\langle A\rangle\langle B\rangle$, where $A$ and $B$ are spin or resonator operators.

Note that in the last equation we introduced a new parameter, $N_\mathrm{S}$, representing the number of spins that are simultaneously coupled to the cavity. In contrast, the second equation that describes the evolution of the spin operator $\sigma_+$ lacks this factor. This reflects the following asymmetry: each spin senses the cavity individually, while the cavity is driven collectively by all spins. In reality, the strength of the driving is different for each spin, and the consequences of this are discussed in the Supplementary Material~\cite{supplementary}. 

In what follows, we will study the steady-state response of the system.  If the longitudinal amplitude $B_\mathrm{l}$ is sufficiently small, we can expand all variables to linear order as
\begin{equation}
    \langle O(t)\rangle \approx\langle O\rangle^{(0)}+\langle O\rangle^{(1)}e^{i\omega_{\mathrm{l}}t}+\langle O\rangle^{(-1)}e^{-i\omega_{\mathrm{l}}t},
\end{equation}
where $O$ represents any of the five operators of the model. Note that the equations are written in the rotating reference frame. If we were to rotate the operators back to the laboratory frame, the transverse spin operators and cavity operators would actually be oscillating at these three frequencies: the central frequency $\omega_{\mathrm{t}}$, the left sideband frequency $\omega_{\mathrm{t}}-\omega_{\mathrm{l}}$, and the right sideband frequency $\omega_{\mathrm{t}}+\omega_{\mathrm{l}}$. As we show in the Supplementary Material~\cite{supplementary}, the sideband emission spectrum is given by $\langle a\rangle^{(\pm 1)}$, up to overall factors.

\section{\label{sec:drivenSpin} Sidebands of a driven spin}

We now turn to a theoretical computation of the sideband response. To set the stage for the full two-dimensional response in the $\Delta_{\mathrm{s}}-\Delta_{\mathrm{c}}$ plane, we begin by working with fixed $\Delta_c$ and analyzing the steady-state features as a function of $\Delta_{\mathrm{s}}$ only. If the spin-cavity feedback is not strong, we can treat the cavity, which acts as a nonlinear band-pass filter, as a source for spin transverse driving having amplitude $B_{\mathrm{t}}$. For given $\Delta_{\mathrm{c}}$, $\gamma_{\mathrm{c}}$, and $K_{\mathrm{c}}$ we determine the steady-state amplitude of this field $|\langle a^{\dagger}\rangle|^2$, by finding the positive solution~\footnote{This is the case when $K_{\mathrm{c}}$ is small, but in general, the cubic equation for $|\langle a^{\dagger}\rangle |^2$ can have three positive solutions (two of which are stable), leading to the bistability of the resonator.} to
\begin{equation}
   (\Delta_{\mathrm{c}}+i\gamma_\mathrm{c})\langle a^{\dagger}\rangle+K_\mathrm{c}|\langle a^{\dagger}\rangle|^2\langle a^{\dagger}\rangle+\frac{g_\mathrm{P}}{2}=0
\end{equation}
which corresponds to setting the left-hand side of Eq.~\eqref{eq:original_systemc} to zero and neglecting the feedback of the spin.

The analysis presented in this section, in one form or another, has appeared elsewhere in the literature~\cite{prior1977, prior1978}. However, for completeness, we derive the main result and examine each regime in detail, in particular, characterizing the regimes relevant to the experiment. Nonlinear response to a strong longitudinal field underpins Landau–Zener interferometry~\cite{shevchenko2010, sergei2019}. An oscillating longitudinal field can be treated as a phase modulation of the transverse drive~\cite{prior1978, camparo1998, camparo2000, camparo2002}, yielding a characteristic Bessel‑function dependence on the ratio of longitudinal amplitude to frequency~\cite{attrash2023}. In that framework, the transverse drive effectively contains infinitely many tones at frequencies $\omega_{\mathrm{t}} + n\omega_{\mathrm{l}}$, where $n$ is an integer, and one typically applies a rotating‑wave approximation (RWA) that retains only the
$k\omega_{\mathrm{l}}$ tone of interest while discarding $n\neq k$ tones. That approximation fails when $B_{\mathrm{t}}$ becomes arbitrarily large, but in the linear regime in $B_{\mathrm{l}}$ we can obtain the full solution in $B_{\mathrm{t}}$ by truncating the Floquet sectors to $k=0,\pm 1$ only.

If we write the effective transverse field as $B_{\mathrm{t}}=\Omega\langle a^{\dagger}\rangle$. We can reduce the number of equations to three,
\begin{eqnarray}
i\frac{d\langle\sigma_+\rangle}{dt} = &-(\Delta_{\mathrm{s}}+B_\mathrm{l}\cos{\left(\omega_\mathrm{l} t\right)}+i\gamma_2) \langle\sigma_{+}\rangle+B_\mathrm{t}\langle\sigma_\mathrm{z}\rangle,\label{eq:spin_system} \\
i\frac{d\langle\sigma_\mathrm{z}\rangle}{dt}= &2\left(B^*_\mathrm{t}\langle\sigma_{+}\rangle-B_{\mathrm{t}}\langle\sigma_{-}\rangle \right)-i\gamma_{1}\left(\langle\sigma_{\mathrm{z}}\rangle-\sigma_{\mathrm{z}}^{\mathrm{eq}}\right),\label{eq:spin_system_p}
\end{eqnarray}
where the equation for $\langle\sigma_-\rangle$ is the complex conjugate of the first equation. When cavity parameters are fixed, emission intensity is determined by $\langle\sigma_+\rangle^{(\pm 1)}$ expectation values, since they are proportional to the cavity expectation values as shown in the Supplementary Material~\cite{supplementary}. We continue the analysis by focusing on the reduced system of equations before going back to the original system to analyze the full two-dimensional spectrum.

The unitary part of the evolution in the reduced system of equations \eqref{eq:spin_system} and \eqref{eq:spin_system_p} can be seen to come from the simpler Hamiltonian, 
\begin{equation}\label{eq:spin_equation}
    H_{\mathrm{s}}^{\mathrm{eff}} = \left(\Delta_{\mathrm{s}}+B_{\mathrm{l}}\cos{\left(\omega_{\mathrm{l}} t\right)}\right) \frac{\sigma_{\mathrm{z}}}{2}+\left(B_{\mathrm{t}}^*\sigma_++B_{\mathrm{t}}\sigma_-\right).
\end{equation}
We are concerned with the limit where the longitudinal field is treated in the linear response regime.

First, we focus on emission at the right sideband frequency $\omega_{\mathrm{t}}+\omega_{\mathrm{l}}$. The intensity of the emission spectrum is then proportional to the square of the absolute value of $\langle\sigma_+\rangle^{(1)}$, as shown in the Supplementary Material~\cite{supplementary}. The full solution for the right sideband, exact in $B_\mathrm{t}$ and expanded to linear order in $B_\mathrm{l}$, is:
\begin{widetext}
\begin{equation}\label{eq:fullequation}
    \langle\sigma_+\rangle^{(1)}=\frac{-B_{\mathrm{l}}B_{\mathrm{t}}\sigma_{\mathrm{z}}^{\text{eq}}}{2\left(\Delta_{\mathrm{s}}-\omega_{\mathrm{l}}+i\gamma_2\right)\left(\Delta_{\mathrm{s}}^2+\gamma_2^2+4\frac{\gamma_2}{\gamma_1}|B_{\mathrm{t}}|^2\right)}\left[\Delta_{\mathrm{s}}-i\gamma_2-\frac{4|B_{\mathrm{t}}|^2\Delta_{\mathrm{s}}\left(\omega_{\mathrm{l}}-2i\gamma_2\right)}{\left(\omega_{\mathrm{l}}-i\gamma_1\right)\left(\Delta_{\mathrm{s}}^2-\left(\omega_{\mathrm{l}}-i\gamma_2\right)^2\right)+4|B_{\mathrm{t}}|^2\left(\omega_{\mathrm{l}}-i\gamma_2\right)}\right],
\end{equation}
\end{widetext}
with the derivation given in Appendix \ref{app:spin}. In what follows, we restrict our attention to situations in which $\omega_{\mathrm{l}}> \gamma_2\gg \gamma_1$, although the expression (\ref{eq:fullequation}) is more generally applicable.

\begin{figure}
    \includegraphics[width=0.49\textwidth]{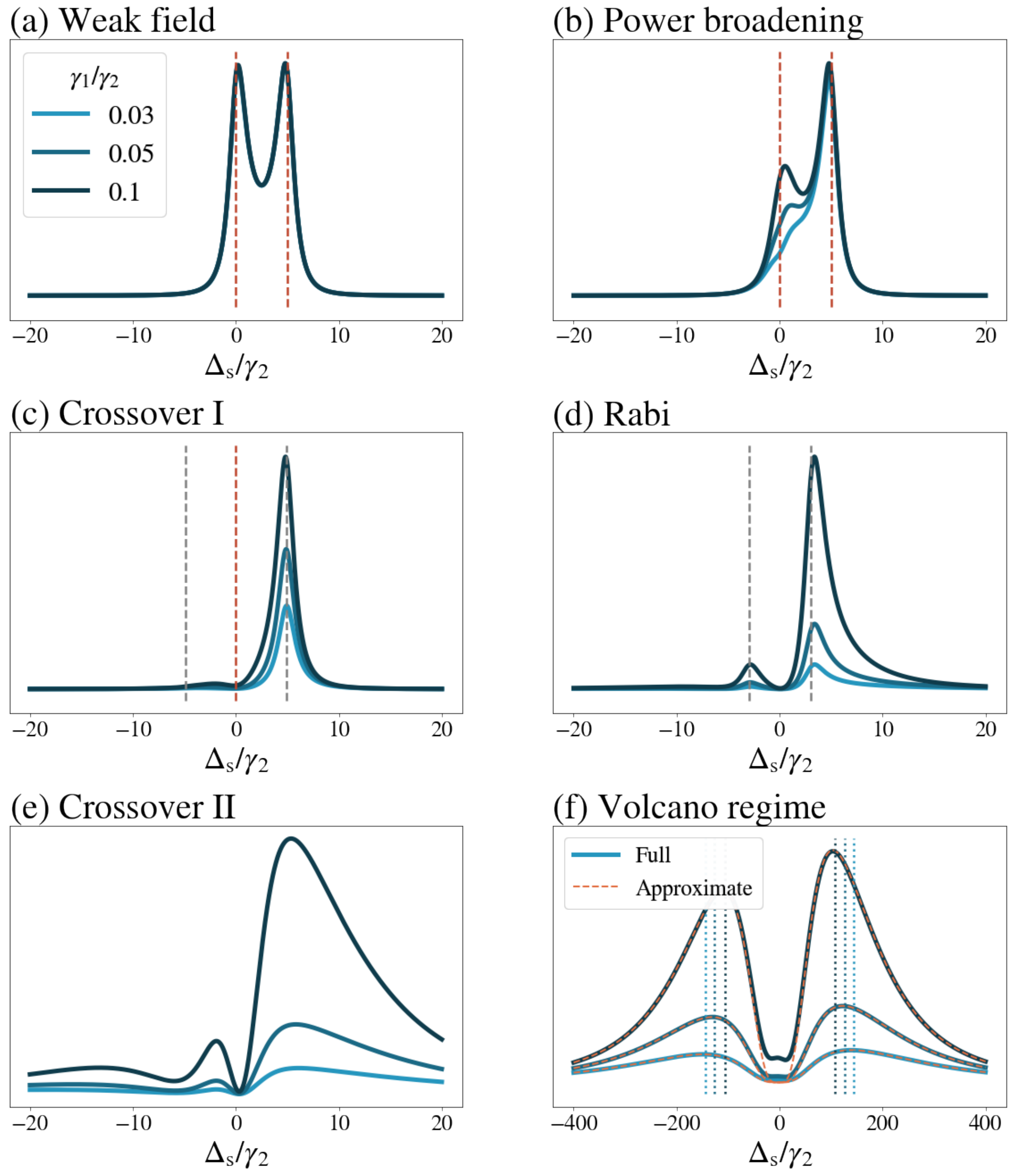}
    \caption{The square of the absolute value of $\langle\sigma_+\rangle^{(1)}$ computed in the $\omega_{\mathrm{l}}\gg\gamma_2\gg\gamma_1$ limit as $B_{\mathrm{t}}$ is gradually increased. Throughout the figure, we keep values $\omega_{\mathrm{l}}=5$, and $\gamma_1$ is varied in each plot with values given in the legend in (a). $B_\mathrm{t}$ goes through values: $0.01$, $0.1$, $0.5$, $2$, $3$, and $30$ from (a) to (f), given in units of $\gamma_2$. The red dashed lines show the positions of $\Delta_{\mathrm{s}}=0$ and $\Delta_{\mathrm{s}}=\omega_{\mathrm{l}}$ resonances, while the gray dashed lines show the positions of the Rabi resonances. In the (f) plot the vertical lines are drawn in the positions of the peaks, i.e. $\pm2|B_{\mathrm{t}}|\sqrt[4]{\frac{\gamma_2}{\gamma_1}}$. Note that this last plot is drawn on a much wider scale than the other plots. The intensity (y-axis) varies significantly between the plots.}
    \label{fig:evolution}
\end{figure}

\begin{figure}
    \includegraphics[width=0.45\textwidth]{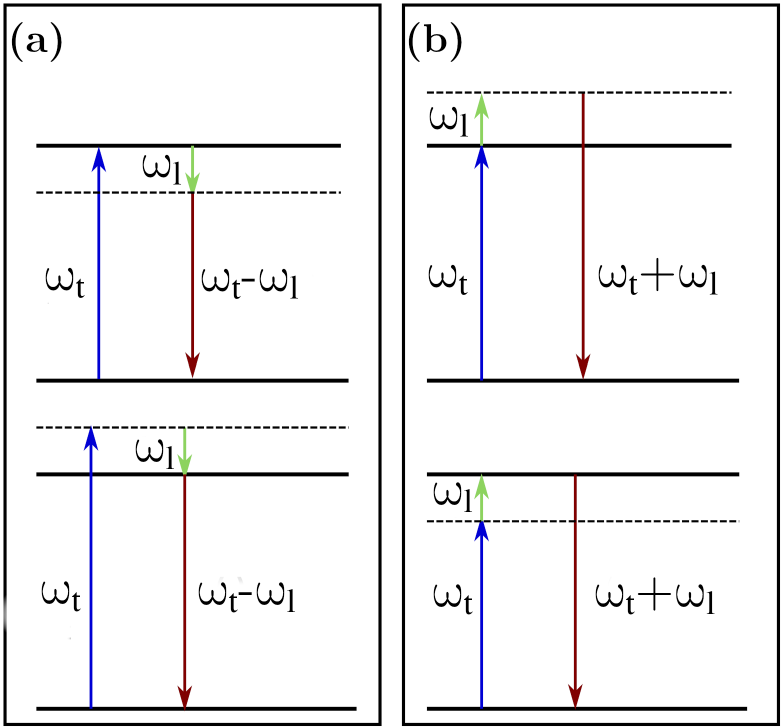}
    \caption{Diagram of the processes that determine the linear response in $B_{\mathrm{t}}$ for (a) the left sideband, and (b) the right sideband.}
    \label{fig:levels}
\end{figure}

The sideband spectrum depends strongly on the transverse drive amplitude $B_{\mathrm{t}}$.  To understand how the spectrum evolves, we begin with the weak-driving limit, when $B_{\mathrm{t}}\ll 1$.  Starting from Eq. (\ref{eq:fullequation}),
the linear response in $B_{\mathrm{t}}$ is easily read-off to be
\begin{equation}
    \langle\sigma_+\rangle^{(1)}=\frac{-B_{\mathrm{l}}B_{\mathrm{t}}\sigma_{\mathrm{z}}^{\text{eq}}}{2\left(\Delta_{\mathrm{s}}-\omega_{\mathrm{l}}+i\gamma_2\right)\left(\Delta_{\mathrm{s}}+i\gamma_2\right)}+\mathcal{O}(B_{\mathrm{t}}^2)\ .
    \label{eq:weakSigma}
\end{equation}
The spectrum has two resonances, at $\Delta_{\mathrm{s}}=0$ and $\Delta_{\mathrm{s}}=\omega_{\mathrm{l}}$, as seen in Fig. \ref{fig:evolution}(a). These arise from the two processes shown in Fig. \ref{fig:levels}(b): the resonance condition is met when the level splitting of the spin, $\omega_\mathrm{s}$, matches either the transverse field frequency, $\omega_\mathrm{t}$, or the sum of frequencies, $\omega_\mathrm{t}+\omega_\mathrm{l}$. 

If we increase the transverse driving the spectrum changes qualitatively. The first effect comes from power broadening, arising from the factor $4\left(\gamma_2/\gamma_1\right) |B_\mathrm{t}|^2$ in Eq.~(\ref{eq:fullequation}).  This factor becomes important at $|B_\mathrm{t}|\approx \frac{1}{2}\sqrt{\gamma_1\gamma_2}$ and it simultaneously broadens and suppresses the resonance at $\Delta_\mathrm{s}=0$, as shown in Fig.~\ref{fig:evolution}(b).

The effects of increasing the driving further are most easily understood from the approximate expression,
\begin{equation}\label{eq:simplified}
   \langle \sigma_+\rangle^{(1)}\approx \frac{-B_{\mathrm{l}}B_{\mathrm{t}}\sigma_{\mathrm{z}}^{\mathrm{eq}}\Delta_{\mathrm{s}}\left(\Delta_{\mathrm{s}}+\omega_{\mathrm{l}}\right)}{2(\Delta_{\mathrm{s}}^2+4\frac{\gamma_2}{\gamma_1}|B_{\mathrm{t}}|^2)(\Delta_{\mathrm{s}}^2+4|B_{\mathrm{t}}|^2-\omega_{\mathrm{l}}^2)},
\end{equation}
which is obtained from Eq.~(\ref{eq:fullequation}) by ignoring the decay rates $\gamma_1$ and $\gamma_2$, but not their ratio $\gamma_2/\gamma_1$.  This expression can be justified in the aforementioned limit
$\omega_{\mathrm{l}}>\gamma_2\gg \gamma_1$, with the caveat that the width of some resonances may need to be regulated by the decay rates.  From this expression, we 
can extract two regimes, depending on the relative magnitude of $|B_\mathrm{t}|$ and $\omega_{\mathrm{l}}$:

\noindent {\it Rabi regime}: When $\sqrt{\gamma_1\gamma_2}\ll 2|\Bt|\ll \oml$, Eq.~(\ref{eq:simplified}) contains two sharp resonances at 
\begin{equation}
\Delta_{\mathrm{s}}=\pm\sqrt{\omega_{\mathrm{l}}^2-4|B_{\mathrm{t}}|^2}.
\label{eq:HHresonance}
\end{equation}
These resonances, shown in Fig.~\ref{fig:evolution}(d) are reached when the longitudinal driving $\oml$ matches the generalized Rabi frequency $\sqrt{\Ds^2+4|\Bt|^2}$.
The width of these resonances is controlled by relaxation rates that appear in the exact expression (\ref{eq:fullequation}).  The two resonances are asymmetric due to the factor of $(\Ds+\oml)$ in the numerator, which suppresses the resonance with the negative sign.

\noindent {\it Volcano regime:}
When $|\Bt|\gg \oml/2$, the condition (\ref{eq:HHresonance}) can no longer be satisfied.  Then, Eq.~(\ref{eq:simplified}) no longer contains any sharp resonances.  Instead, a volcano-shaped spectrum emerges, as shown in Fig.~\ref{fig:evolution}(f), reflecting the interplay between broad resonances centered at $\Ds=0$ and the factor $\Ds(\Ds+\oml)$ in the numerator, which creates a transparency region near $\Ds=0$.  The width of the volcano is of order 
$2\sqrt{\gamma_2/\gamma_1}|\Bt|$, while the width of the central transparency region is $2|\Bt|$.  As in the Rabi regime, the factor of $(\Ds+\oml)$ gives rise to asymmetry between the two peaks of the volcano.  The location of the peaks of the volcano can be found from the local maxima of Eq. \eqref{eq:simplified}. This leads  to the quintic equation from which the peak positions can be estimated, and in the limit $\omega_{\mathrm{l}}\ll2|B_{\mathrm{t}}|$ they are 
\begin{equation}
        \Delta_{\mathrm{max}} \approx \pm 2|B_{\mathrm{t}}|\sqrt[4]{\frac{\gamma_2}{\gamma_1}}-\frac{\omega_{\mathrm{l}}}{8}\left(1+\sqrt{\frac{\gamma_1}{\gamma_2}}\right)+\mathcal{O}\left(\frac{\omega_{\mathrm{l}}}{|B_{\mathrm{t}}|^2}\right),
\end{equation}
where $\Delta_{\mathrm{max}}$ is the detuning at which two broad peaks reach their maximum intensity.

In addition, the spectrum displays crossovers at the interface between the different regimes.  At the crossover between weak-field and Rabi regimes [Fig.~\ref{fig:evolution}(c)], for $\sqrt{\gamma_1\gamma_2}\ll 2|\Bt|\lesssim \gamma_2$, only the right Rabi resonance is visible. On the other hand, at the crossover between the Rabi and volcano regimes [Fig.~\ref{fig:evolution}(e)], for $2|\Bt|\approx \oml$, the Rabi peaks are not sharp, but they have not been sufficiently broadened for the volcano structure to emerge.  We call these crossovers I and II, respectively. The crossover II between the Rabi and the volcano regime is illustrated through the analysis of the motion of complex poles of the response function in App.~\ref{app:complex_poles}.

\begin{figure}
    \centering
    \includegraphics[width=.49\textwidth]{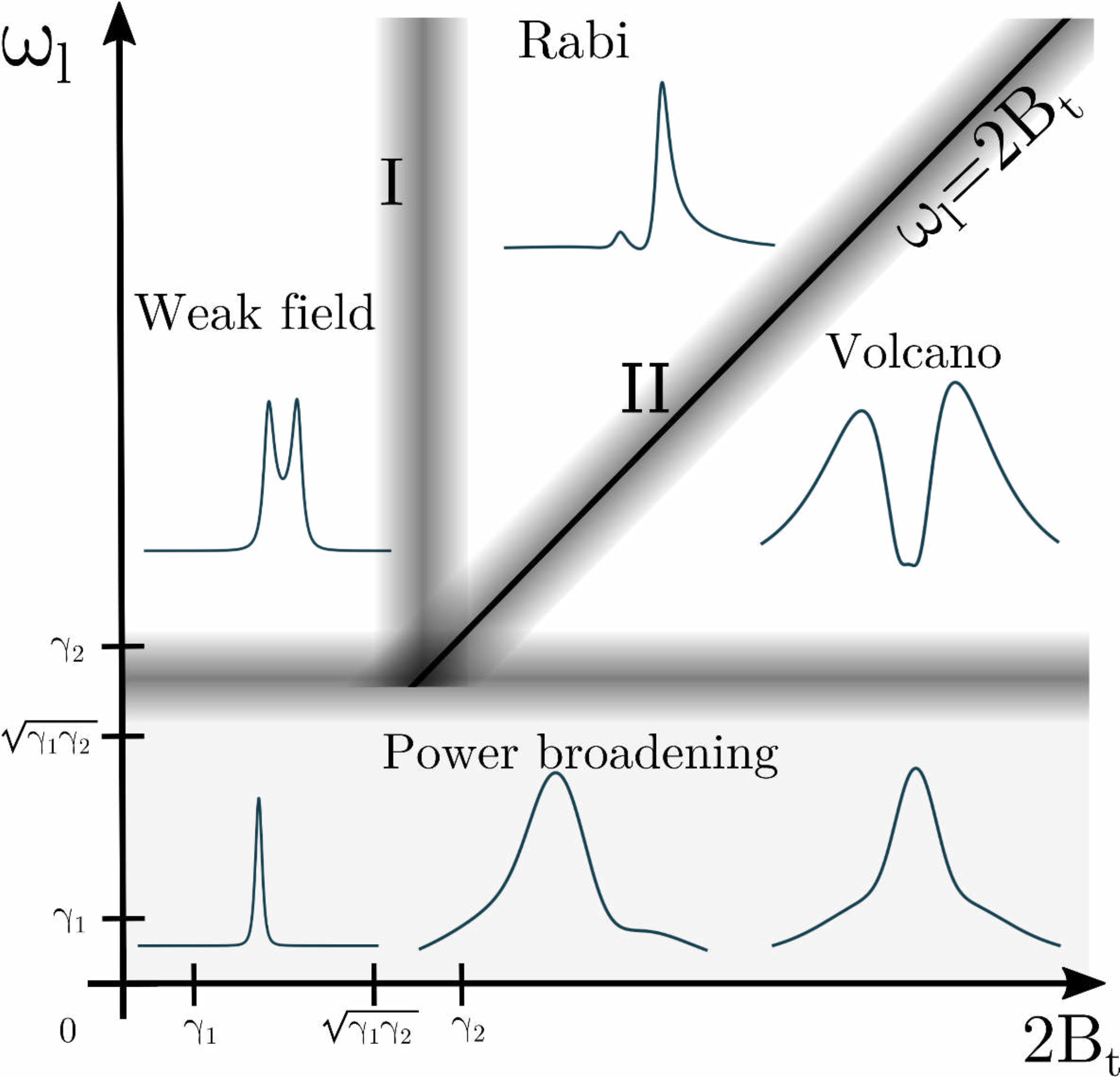}
    \caption{Different regimes of the sideband response. When $\omega_{\mathrm{l}}>\gamma_2$, as $B_{\mathrm{t}}$ is increased we encounter three different regimes separated by crossover regions $\mathrm{I}$ and $\mathrm{II}$ that are shaded in the figure. The crossover $\mathrm{II}$ between the Rabi and the volcano regimes lies around the $\omega_{\mathrm{l}}=2B_{\mathrm{t}}$ solid line. In the slow field limit ($\omega_{\mathrm{l}}<\gamma_2$), the spectrum evolves smoothly without a clear separation between different regimes.}
    \label{fig:regimes}
\end{figure}

Figure \ref{fig:regimes} shows the different regimes in the $\Bt$-$\oml$ plane.  Here, we also show the spectrum in the slow longitudinal field limit $\oml<\gamma_2$, which is generically seen to only contain a single peak regardless of the value of $\Bt$ (except for certain fine-tuned parameters, where a richer structure can be obtained).  The width of this peak tends to increase with $\Bt$ due to power broadening.

This broadening, which comes from the $B_{\mathrm{t}}$--dependent resonance that multiplies brackets in expression \eqref{eq:fullequation} is a common obstacle in spectroscopy. It also appears in the central frequency response, as it can be seen from Eq.~\eqref{eq:occupation_zero}, where it leads to a featureless spectrum. However, as we showed, in the sideband response, when $\omega_{\mathrm{l}}$ is larger than $\gamma_2$, the spectrum develops many interesting features that are not limited by power broadening. In the other limit, if we let $\omega_{\mathrm{l}}$ tend to $0$, we recover the central frequency power-broadened response. When $\omega_{\mathrm{l}}$ is very small but finite, most of the changes to the spectrum will still be washed out by this broadening.

Thus far, we have focused exclusively on the right sideband response.  We now turn our attention to the left sideband, whose spectrum is given by the absolute value of $\langle\sigma_+\rangle^{(-1)}$. In the reduced model that we use in this section, $\langle\sigma_+\rangle^{(-1)}$ is simply obtained by replacing $\oml$ with $-\oml$ in the expression for $\langle\sigma_+\rangle^{(1)}$ in Eq.~(\ref{eq:fullequation}). This replacement is equivalent to the transformation that consists of complex conjugation combined with the replacement of $\Delta_{\mathrm{s}}$ by $-\Delta_{\mathrm{s}}$. Since the spectrum only depends on the absolute value, the complex conjugation can be ignored.  Hence, the left sideband spectrum is the mirror image of the right sideband spectrum, under the reflection $\Ds\to-\Ds$.

\section{\label{sec:spin_cavity}Spin-cavity sidebands}

We return to the original problem of a coupled spin-cavity system [see Sec.~\ref{sec:setup}], with two external drives: the spin is driven by the longitudinal field at frequency $\omega_{\mathrm{l}}$ and the cavity is driven at frequency $\omega_{\mathrm{t}}$. The cavity transfers oscillations to the spin at the latter frequency through an effective magnetic field $B_{\mathrm{t}}=\Omega\langle a^{\dagger}\rangle^{(0)}$. This field is not treated as an external parameter, as in Sec.~\ref{sec:drivenSpin}, but it is determined by a single-mode driven nonlinear cavity with natural frequency $\omega_{\mathrm{c}}$. Since the spin sees two drivings, an external driving at frequency $\omega_{\mathrm{l}}$ and an effective cavity driving at frequency $\omega_{\mathrm{t}}$, it can act as a frequency-mixer to produce radiation at the sum and the difference of these two frequencies. This sideband radiation, in turn, drives the cavity at $\omega_{\mathrm{t}}\pm\omega_{\mathrm{l}}$, and we can read out the sideband response of the spin from the cavity oscillations. 

The simplified description that we have given ignores the higher-order spin-cavity feedback effects. In actuality, the effective cavity driving at $\omega_{\mathrm{t}}$ is renormalized by the spin oscillations themselves.  However, the effect of this feedback on the sideband spectrum is negligible if the spin-cavity coupling strength is weak. In addition to this, we will assume at first that the cavity is linear. We use this as a basis from which we will later show how making the cavity nonlinear brings about novel effects.

\begin{figure}
    \centering
    \includegraphics[width=.49\textwidth]{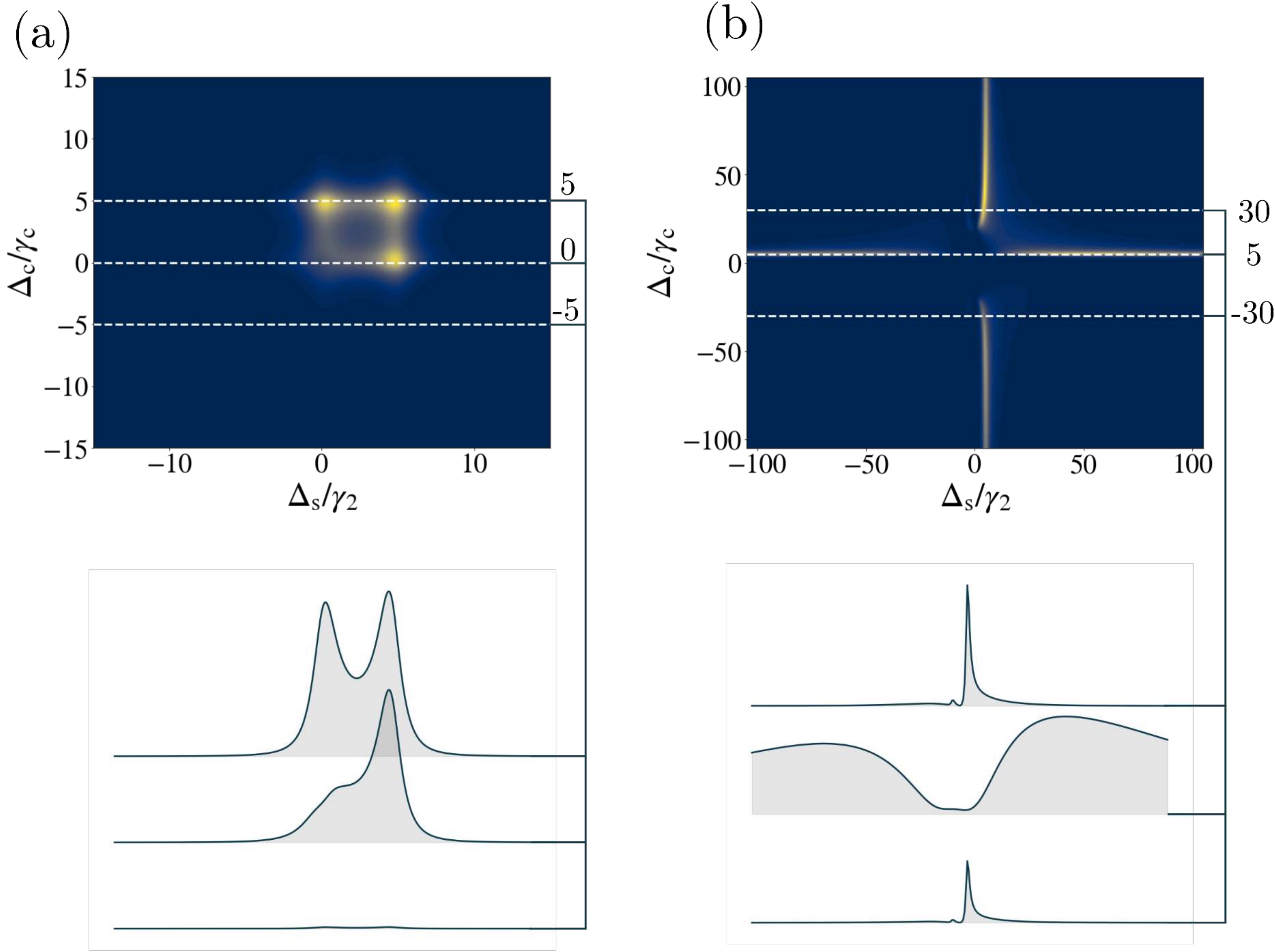}
    \caption{$2$D right sideband response given by $|\langle a^{\dagger}\rangle^{( 1)}|^2$ without the cavity nonlinearity and the spin feedback. The strength of the cavity driving is weak in (a) where $g_{\mathrm{P}}$ is $0.1$, and strong in (b), where $g_{\mathrm{P}}=100.0$. The slices along which the one-dimensional response is shown are indicated by the horizontal dashed lines. They are at the positions $\Delta_{\mathrm{c}}=0,\hspace{1mm} \pm \omega_{\mathrm{l}}$ in (a). The signal is amplified when $\Delta_{\mathrm{c}}=\omega_{\mathrm{l}}$ for the right sideband and this fact is reflected in the plots. However, when $\Delta_{\mathrm{c}}=0$ the spin driving is the strongest and near the linear response regime (a) we see the signal at this point. The strong field leads to power broadening and this is the reason why we see only one of two resonances. In (b) we see the transition from the volcano to the Rabi regime as $\Dc$ is varied. The volcano spectrum is enhanced near $\Delta_{\mathrm{c}}=\omega_{\mathrm{l}}$. Even though the Rabi regime is far from the amplification region, we can still see the peaks because they are much sharper than the broad volcano. The rest of the parameters of the plots are given in units of $\gamma_2$: $\omega_{\mathrm{l}}=5.0$, $\gamma_1=0.01$, $\gamma_{\mathrm{c}}=1.0$, and $\Omega=0.01$.}
    \label{fig:sc_evolution}
\end{figure}

\subsection{The Right Sideband}

We now study the right sideband spectral intensity function in the two-dimensional $\Ds-\Dc$ plane. In other words, in Sec.~\ref{sec:drivenSpin} we presented the $\Delta_{\mathrm{s}}$-dependence of the response by fixing the cavity parameters, and now we want to study the $\Dc$-dependence as well.  In the Supplementary Material~\cite{supplementary} we show that the intensity of the right sideband is determined by the $|\langle a^{\dagger}\rangle^{(1)}|^2$ expectation value.   Within the limits mentioned earlier, in which the Kerr nonlinearity and feedback effects are neglected, the steady state solution for the cavity equation (Eq. \ref{eq:original_systemc}) reduces to the damped Lorentzian oscillator solution, $\langle a^{\dagger}\rangle^{(0)}=\frac{-g_{\mathrm{P}}}{2(\Dc+i\gamma_{\mathrm{c}})}$. Therefore, the effective transverse field that each spin feels is
\begin{equation}
B_{\mathrm{t}}=\frac{-g_{\mathrm{P}}\Omega}{2(\Dc+i\gamma_{\mathrm{c}})}
\label{eq:BtEff}
\end{equation}
The amplitude of the field is highest when the external driving is resonant with the cavity ($\Delta_{\mathrm{c}}=\omega_{\mathrm{c}}-\omega_{\mathrm{t}}=0$) and its value is $\frac{g_{\mathrm{P}}\Omega}{2\gamma_{\mathrm{c}}}$. As the external driving is detuned from the cavity, the amplitude falls off with the width set by the decay rate $\gamma_{\mathrm{c}}$. The right sideband steady state solution for the cavity is then
\begin{equation}
    \langle a^{\dagger}\rangle^{(1)}=-\frac{N_\mathrm{S}\Omega}{\Delta_{\mathrm{c}}-\omega_{\mathrm{l}}+i\gamma_{\mathrm{c}}}\langle\sigma_{+}\rangle^{(1)}(B_{\mathrm{t}})\Bigg|_{B_{\mathrm{t}}=\frac{-g_{\mathrm{P}}\Omega}{2(\Dc+i\gamma_{\mathrm{c}})}},
    \label{eq:simplified_cavity}
\end{equation}
with the derivation given in Appendix \ref{app:spin_cavity}. 

Equation (\ref{eq:simplified_cavity}) reflects the twofold role of the cavity on the sideband response. First, the cavity provides an effective transverse drive to the spins whose amplitude is given in Eq. (\ref{eq:BtEff}). This drive is strongest when the cavity is at resonance with the transverse frequency $\omega_\mathrm{t}$. Second, the cavity acts as an antenna that measures the response itself.  This gives rise to the amplification factor in front of Eq. (\ref{eq:simplified_cavity}), which is maximal when the cavity is at resonance with the sideband frequency $\omega_\mathrm{t}+\omega_\mathrm{l}$.  While Eq.~(\ref{eq:simplified_cavity}) ignores feedback effects, we have checked explicitly that it captures the full solution to the mean field equations (\ref{eq:full_system}) to an excellent approximation whenever the cavity-spin coupling is small.

When the cavity is weakly driven ($g_{\mathrm{P}}\ll 1$) we obtain from the linear response solution in Eq. (\ref{eq:weakSigma}), 
\begin{equation}
\frac{-g_{\mathrm{P}}N_{\mathrm{S}}\Omega^2\sigma_{\mathrm{z}}^{\text{eq}}}{4\left(\Delta_{\mathrm{s}}-\omega_{\mathrm{l}}+i\gamma_2\right)\left(\Delta_{\mathrm{s}}+i\gamma_2\right)\left(\Delta_{\mathrm{c}}+i\gamma_{\mathrm{c}}\right)\left(\Delta_\mathrm{c}-\omega_{\mathrm{l}}+i\gamma_{\mathrm{c}}\right)}\ .
\end{equation}
This response shows four resonances.  The two resonances at $\Ds=0$ and $\Ds=\oml$ were discussed in Sec.~\ref{sec:drivenSpin}.  The resonance at $\Delta_{\mathrm{c}}=0$ comes from the fact that near the linear response regime, the spectral intensity is proportional to the magnitude of $B_{\mathrm{t}}$, which assumes its maximum value near $\Delta_{\mathrm{c}}=0$. The last resonance occurs when $\Delta_{\mathrm{c}}=\omega_{\mathrm{l}}$, and it describes the amplification of the signal by the cavity.

If the cavity driving strength $g_\mathrm{P}$ is increased, the effective magnetic field felt by the spin increases, and power broadening suppresses the $\Delta_{\mathrm{s}}=0$ resonance. Since the magnetic field is highest near the $\Delta_{\mathrm{c}}=0$ resonance we expect this suppression to occur first near this resonance. This scenario is shown in Fig. \ref{fig:sc_evolution}(a). Out of the four resonances in the $\Delta_{\mathrm{s}}-\Delta_{\mathrm{c}}$ plane, the first one to be suppressed is at $\Delta_{\mathrm{s}}=\Delta_{\mathrm{c}}=0$.  

After an additional increase in cavity driving strength we obtain the spectrum shown in Fig.~\ref{fig:sc_evolution}(b).  This figure demonstrates the transition between the Rabi and the volcano regimes, which can be seen for a fixed set of parameters. Near $\Delta_{\mathrm{c}}=\omega_{\mathrm{l}}$ the effective field amplitude (\ref{eq:BtEff}) is large ($|B_{\mathrm{t}}|\gg\omega_{\mathrm{l}}$), corresponding to the volcano regime. Although the spectral weight of the volcano is spread out, the signal is amplified by the cavity and it is therefore clearly visible in the spectrum. On the other hand, for large $\Delta_{\mathrm{c}}$ the amplification of the sideband radiation is much weaker, and one would expect not to see significant response. However, when $\omega_{\mathrm{l}}\gg2|B_{\mathrm{t}}|$ we enter the Rabi regime, in which the resonances are very sharp.  This leads to visible peaks even though the amplification is weak. Which of these two features, volcano and Rabi, is more visible depends on the relative sizes of $\gamma_2$ and $\gamma_{\mathrm{c}}$: $\gamma_{2}$ sets the width of the Rabi resonances, while $\gamma_{\mathrm{c}}$ sets the width of the amplification region. 

\subsection{The Sideband Symmetry}

\label{subsec:SidebandSymmetry}

\begin{figure}
    \centering
    \includegraphics[width=.48\textwidth]{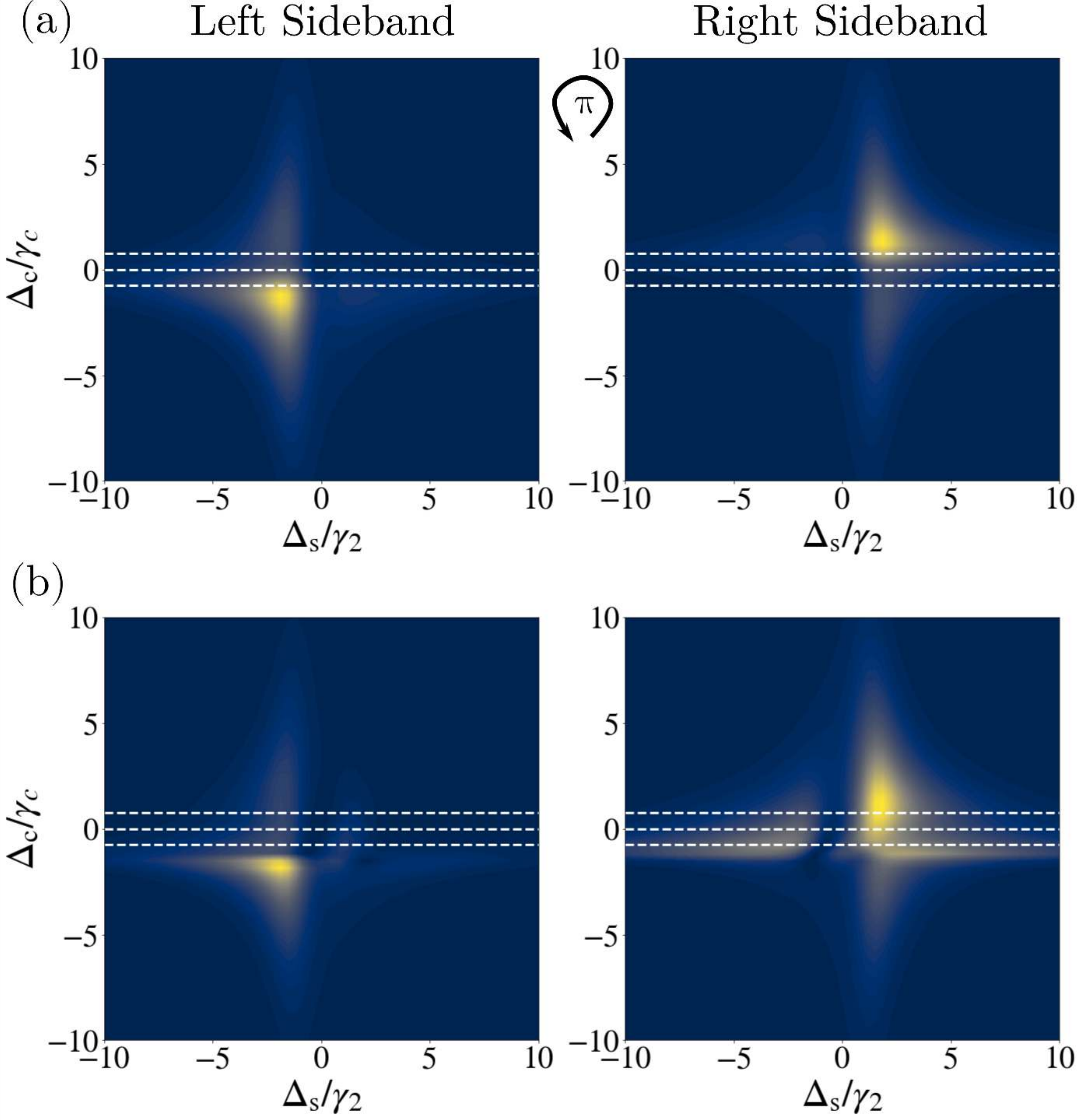}
    \caption{Comparison between the $2$D-spectra of the theoretically computed left and the right sideband. (a) In the absence of nonlinearity, when $\eta=0$, the right sideband spectrum is obtained by $\pi$-rotation ($\Delta_{\mathrm{c}}\rightarrow-\Delta_{\mathrm{c}}$, and $\Delta_{\mathrm{s}}\rightarrow-\Delta_{\mathrm{s}}$) of the left sideband spectrum. (b) When Kerr nonlinearity is added ($\eta=1$), the symmetry between sidebands is broken. The parameters in units of $\gamma_2$ are $\omega_{\mathrm{l}}=1.5,\ \gamma_1=0.01, \gamma_{\mathrm{c}}=2.0, \ g_{\mathrm{P}}=200.0, \ \Omega=0.01, \ N_{\mathrm{S}}=1,\ \mathrm{and}\ \sigma_{\mathrm{z}}^{\mathrm{eq}}=-1$.}
    \label{fig:symmetry}
\end{figure}

In Sec.~\ref{sec:drivenSpin} we showed that the left sideband response can be obtained from the right sideband by replacing $\Delta_{\mathrm{s}}$ with $-\Ds$, but keeping $\Bt$ constant. A natural question to ask is whether this symmetry extends to the full two-dimensional spectrum.  In App.~\ref{app:spin_cavity} we show that the symmetry condition can be generalized: the left sideband is obtained from the right sideband by simultaneously replacing $\Ds$ with $-\Ds$ and $\Dc$ with $-\Dc$. This can be seen as a composition of two reflections in the $\Ds-\Dc$ plane or, equivalently, as a $\pi$-angle rotation of the plane. The symmetry is present even when the spin-cavity feedback is not ignored. One consequence of the symmetry is that the left sideband is most strongly amplified near $\Dc=-\omega_{\mathrm{l}}$. The relation between the left and the right sideband spectra is illustrated in Fig. \ref{fig:symmetry}(a).

This symmetry relation is clearly violated in the experimentally obtained spectra shown in the bottom row of Fig.~\ref{fig:experimental_volcano}. The general symmetry-based principle excludes a large class of linear models and points to a nonlinear origin of the asymmetry. Although the exact details of the experimental conditions in the complex diamond-defect setup are not known, we illustrate here how Hamiltonian nonlinearity can give rise to this counterintuitive response. By introducing Kerr nonlinearity to the cavity subsystem, we break the symmetry relation. This nonlinearity enables the cavity to mix frequencies in an asymmetric way between the left and the right sidebands. The breaking of the symmetry relation between sidebands is shown in Fig. \ref{fig:symmetry}(b). We will next analyze the effects arising from such cavity nonlinearity. 

\section{\label{sec:nonlinear_cavity} Cavity Nonlinearity and The Sideband Anomaly}

Nonlinearity of the resonator gives rise to an amplitude-dependent shift of its resonant frequency. As an example, the Duffing resonator with a cubic nonlinear term in the equations of motion exhibits a blue shift for positive nonlinearity coefficient and a red shift for negative coefficient. The Kerr nonlinearity of the cavity similarly gives rise to the cubic term in the optical Bloch equations. However, the nonlinearity can also mix frequencies and consequently modify the emission spectrum. 

We show that if the resonator is driven to its nonlinear regime, the symmetry between the two sidebands is broken. We demonstrate our results with the example of the Kerr nonlinearity. First, we recall that the nonlinearity leads to the blue-shifted tilted central spectral line in the response (when the Kerr coefficient  $K_{\mathrm{c}}$ is positive). We show that the influence on the sidebands is more complicated. The magnitude of the blueshift is a non-monotonic function of the longitudinal frequency and the spectral response can exhibit many complicated features. This hypersensitivity of the nonlinear response to the physical parameters points to its usefulness as a gauge of the internal dynamics of quantum systems.

The cavity is a dynamical system that both drives and measures the spins. The effects of the Kerr nonlinearity on the response are mediated through the spin-cavity feedback and become particularly visible in the sidebands. Throughout the section, we treat the nonlinearity and the spin-cavity feedback perturbatively. The perturbative solution restricts the response to be outside of the bistability region, which requires a strongly excited nonlinearity. This is the region with the two stable steady-state solutions for the Duffing resonator.

The Kerr nonlinearity with $K_{\mathrm{c}}>0$ blueshifts the cavity frequency. This can be seen in Fig.~\ref{fig:combined_center}(c), where the spectral line is slanted to the left. As shown in App.~\ref{app:central}, parameter
\begin{equation}
\eta = \frac{g_\mathrm{P}^2K_\mathrm{c}}{4\gamma_\mathrm{c}^3},
\end{equation}
determines the strength of the Kerr nonlinearity. To the cubic order in $\eta$, and away from the spin resonance where the spin-cavity feedback becomes prominent, the position of the maximum of the central resonance is shifted to
\begin{equation}\label{eq:center_shift}
    \Delta_{\mathrm{c,\;res}}^{(0)}|_{\Delta_\mathrm{s}\rightarrow\infty}=-\eta\gamma_{\mathrm{c}}+\mathcal{O}\left(\eta^3\right).
\end{equation}
This shift of the resonant detuning to the left corresponds to the blue shift of the cavity frequency. As $\eta$ is varied in panel (c) of Fig.~\ref{fig:combined_center}, the maximum of the resonance closely follows this simple approximate result. It is worth noting that a further increase of $\eta$ would lead the cavity to the optical bistability region.

\begin{figure*}
\centering\includegraphics[width=0.65\textwidth]{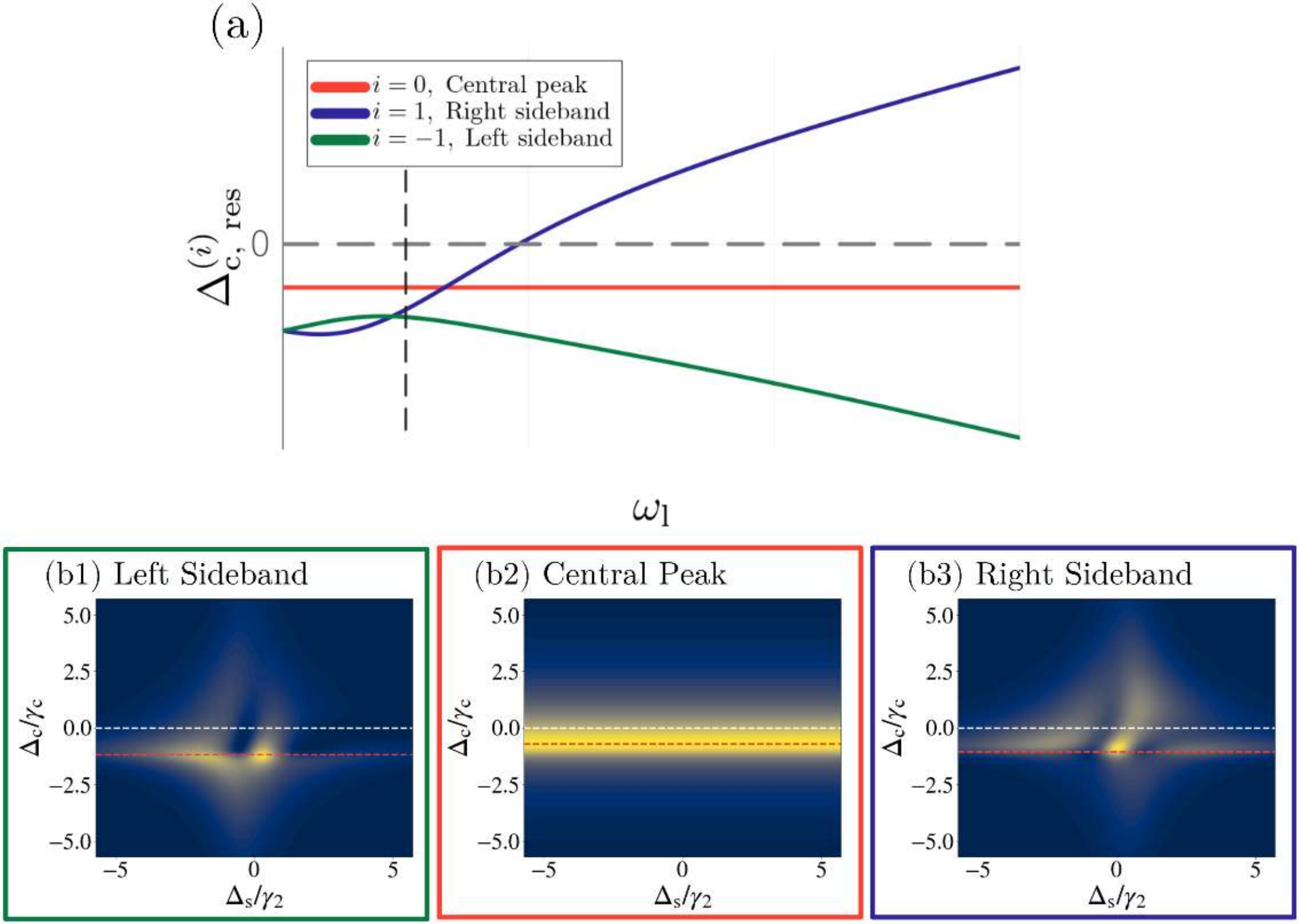}    \caption{The positions of the central resonance, $\Delta_\mathrm{c,\ res}^{(0)}$, and the sideband resonances, $\Delta_\mathrm{c,\ res}^{(\pm 1)}$, depending on the frequency of the longitudinal field $\omega_{\mathrm{l}}$ are shown in (a). The strength of the Kerr nonlinearity is $\eta=0.7$, and that is enough to lead to an anomalous behavior when $\omega_{\mathrm{l}}$ is comparable to $\gamma_{\mathrm{c}}$. The $2$D spectra for the central line and the sidebands (b1-3) are drawn when $\omega_{\mathrm{l}}=0.5\gamma_{\mathrm{c}}$. In this region, both of the sidebands are shifted to the left of the central resonance. The rest of the parameters of the plots are $g_{\mathrm{P}}=50.0, \ \Omega=0.01, \ N_{\mathrm{S}}=1,\ \sigma_{\mathrm{z}}^{\mathrm{eq}}=-1.0, \ \gamma_1 = 0.01,\ \mathrm{and }\ \gamma_{\mathrm{c}} = 1.0$.}
    \label{fig:two_effect}
\end{figure*}

The effect of the Kerr nonlinearity on sidebands is subtler, and it is treated in detail in App.~\ref{app:sidebands}. However, the most direct effect is the shift of the spectral lines. We focus on the case when $K_{\mathrm{c}}>0$, which leads to the blue shift of the spectrum. The results can be straightforwardly generalized to the case of the negative Kerr coefficient $(K_{\mathrm{c}}<0)$, though this requires higher-order terms to stabilize the Hamiltonian. As it is shown in Appendix \ref{app:spin_cavity}, the resonances that govern the amplification of the sideband signals appear. The positions of the resonance centers are to cubic order in $\eta$,

\begin{equation}\label{eq:sideband_shift}
\begin{split}
    \Delta_{\mathrm{c,\; res}}^{(\pm 1)}&=\pm\omega_\mathrm{l}-2\eta\frac{\gamma_{\mathrm{c}}}{1+\left(\frac{\omega_\mathrm{l}}{\gamma_\mathrm{c}}\right)^2}\\&\mp\frac{7}{2}\eta^2\frac{\omega_{\mathrm{l}}}{\left(1+\left(\frac{\omega_\mathrm{l}}{\gamma_\mathrm{c}}\right)^2\right)^3}+\mathcal{O}\left(\eta^3\right).
\end{split}
\end{equation}
The positions of the central resonance and the two sidebands as functions of the longitudinal frequency $\omega_{\mathrm{l}}$ are shown in Fig.~\ref{fig:two_effect}(a).

The first term in Eq.~\eqref{eq:sideband_shift} exists in the absence of the Kerr nonlinearity ($\eta=0$). It is a simple condition $\Delta_{\mathrm{c}}=\omega_{\mathrm{c}}-\omega_{\mathrm{t}}=\pm\omega_\mathrm{l}$ for the sideband light to hit the cavity resonance. The second term creates the blue shift when the longitudinal frequency is small compared to the cavity width $\gamma_{\mathrm{c}}$.  Importantly, it shifts the right and the left sidebands in the same direction. When $\omega_{\mathrm{l}}\ll \gamma_c$ this shift is dominant and it comes with a non-trivial factor of $2$. This factor is a consequence of the frequency mixing that occurs due to the nonlinear cavity term $K_{\mathrm{c}} a^{\dagger}a^{\dagger}a$ from Eq.~\eqref{eq:original_systemc}. As $\omega_{\mathrm{l}}$ is increased significantly beyond the cavity width $\gamma_{\mathrm{c}}$, this effect becomes less significant since the off-resonant sidebands cannot excite the cavity nonlinearity strongly. In the limit when $\omega_{\mathrm{l}}\gg\gamma_{\mathrm{c}}$ the position of the resonance can be primarily understood as a simple resonant amplification of sideband radiation if the sideband frequency matches the bare cavity frequency $\omega_{\mathrm{c}}$. The intermediate region is resolved through an interplay of \emph{the resonant sideband amplification} when the sideband photons hit the cavity resonance and \emph{the excitation of the Kerr nonlinearity}, which leads to frequency mixing.

It can be read out from expression~\eqref{eq:sideband_shift} that when $\omega_{\mathrm{l}}\ll \gamma_c$ the shift of the resonance of both of the sidebands is twice the shift of the central resonance (Eq.~\eqref{eq:center_shift}). The right sideband is to the left of the resonance, and the \emph{sideband anomaly} always occurs at very slow longitudinal fields. However, when $\omega_{\mathrm{l}}$ is sufficiently small the sideband signals are indiscernible because of the broadening effects. Therefore, we look at the condition for the anomaly at a finite $\omega_{\mathrm{l}}$. We define the nonlinearity strength $\eta_{\mathrm{anom}}\left(\omega_{\mathrm{l}}/\gamma_{\mathrm{c}}\right)$ as the condition for the onset of the anomaly,
\begin{equation}
    \Delta_{\mathrm{c,\; res}}^{( 1)}\left[\eta_{\mathrm{anom}}\left(\frac{\omega_{\mathrm{l}}}{\gamma_{\mathrm{c}}}\right)\right]=\Delta_{\mathrm{c,\; res}}^{(0)}\left[\eta_{\mathrm{anom}}\left(\frac{\omega_{\mathrm{l}}}{\gamma_{\mathrm{c}}}\right)\right].
\end{equation}
The solution of the equation for $\eta_{\mathrm{anom}}=\frac{g_{\mathrm{P}}^2}{4\gamma_{\mathrm{c}}^3}K_{\mathrm{c}}^{\mathrm{anom}}$ gives the anomaly Kerr strength $K_{\mathrm{c}}^{\mathrm{anom}}$ for the onset of the sideband anomaly,
\begin{align}
\label{eq:critical_condition}
\nonumber&K_{\mathrm{c}}^{\mathrm{anom}}=\frac{4}{7g_{\mathrm{P}}^{2}}\Bigg[\frac{\left(\omega_{\mathrm{l}}^2-\gamma_{\mathrm{c}}^2\right)\left(\omega_{\mathrm{l}}^2+\gamma_{\mathrm{c}}^2\right)^2}{\omega_{\mathrm{l}}\gamma_{\mathrm{c}}^2}\\ &+\sqrt{\frac{\left(\omega_{\mathrm{l}}^2-\gamma_{\mathrm{c}}^2\right)^2\left(\omega_{\mathrm{l}}^2+\gamma_{\mathrm{c}}^2\right)^4}{\omega_{\mathrm{l}}^2\gamma_{\mathrm{c}}^4}+14\left(\omega_{\mathrm{l}}^2+\gamma_{\mathrm{c}}^2\right)^3}\Bigg]+\mathcal{O}\left(\eta^3\right).
\end{align}

In Figure~\ref{fig:two_effect}(a), the nonlinearity strength $\eta$ is large enough that the order of the right and the left sidebands becomes inverted at low $\omega_{\mathrm{l}}$. This effect comes from the quadratic term in Eq.~\eqref{eq:sideband_shift}. One should be careful when considering these analytic results; the nonlinearity strength becomes large enough to render the approximate results qualitative. With this cautionary remark, we write the expression for the distance between the sidebands,    
\begin{equation}\label{eq:resonance_difference}
\begin{split}
    \Delta_{\mathrm{c,\; res}}^{( 1)}-\Delta_{\mathrm{c,\; res}}^{( -1)}&=2\omega_\mathrm{l}\left(1-\frac{7g_{\mathrm{P}}^4K_{\mathrm{c}}^2}{32\left(\omega_{\mathrm{l}}^2+\gamma_{\mathrm{c}}^2\right)^3}\right)+\mathcal{O}\left(\eta^3\right).
\end{split}
\end{equation}
When the term in the brackets vanishes, the order of the left and the right sideband is inverted. This happens at the inversion Kerr strength, 
\begin{equation}\label{eq:inversion_condition}
    K_{\mathrm{c}}^{\mathrm{inv}}=\frac{4}{g_{\mathrm{P}}^2}\sqrt{\frac{2\left(\omega_{\mathrm{l}}^2+\gamma_{\mathrm{c}}^2\right)^3}{7}}+\mathcal{O}\left(\eta^3\right).
\end{equation}

\begin{figure}
    \centering
    \includegraphics[width=0.49\textwidth]{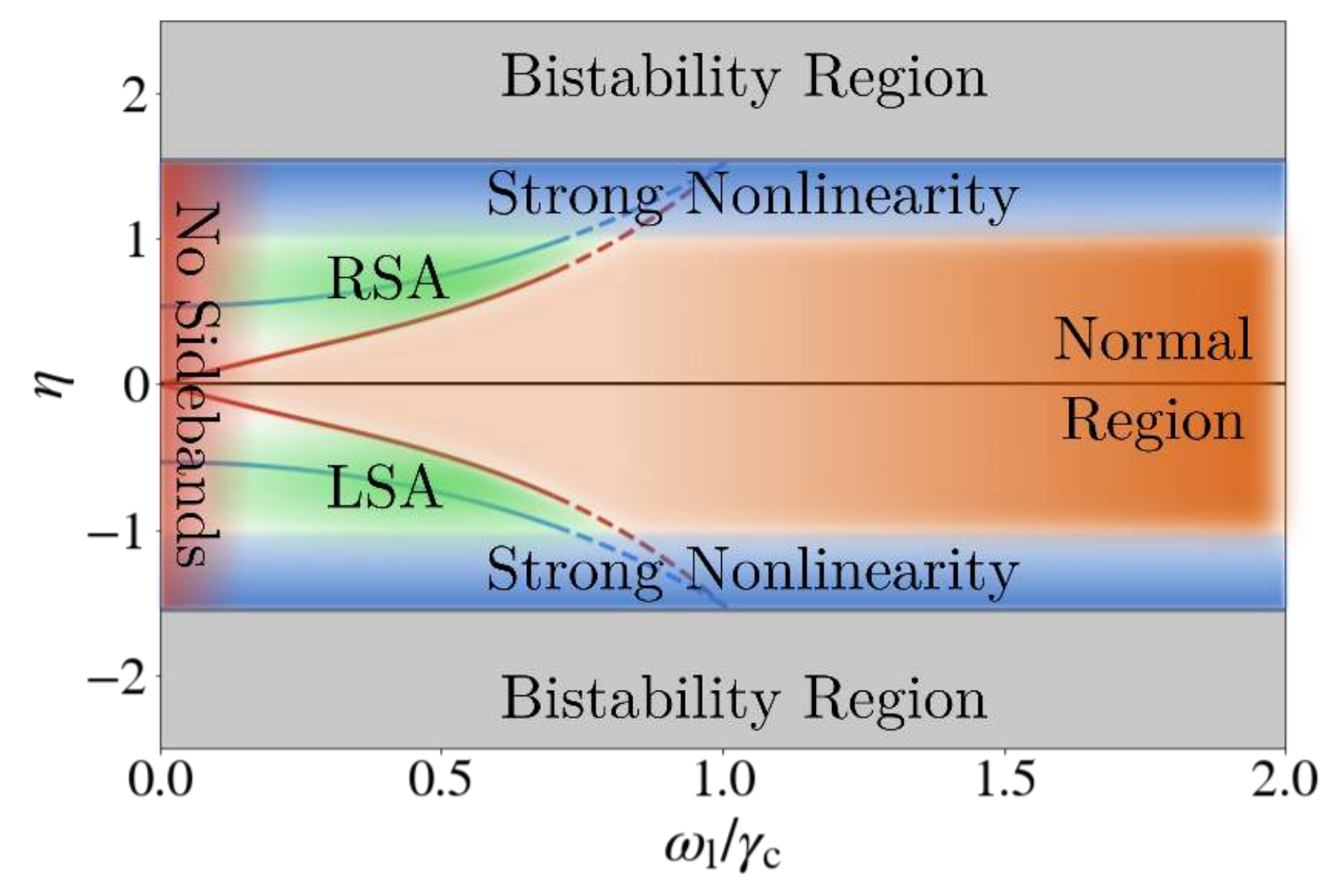}
    \caption{The regions of the right-sideband anomaly (RSA) and the left-sideband anomaly (LSA) shown in the diagram with the nonlinearity strength $\eta$ on the $y$-axis and the normalized longitudinal frequency $\omega_{\mathrm{l}}/\gamma_{\mathrm{c}}$ on the $x$-axis. The line separating two anomalous regions from the region without the anomaly (normal region) is calculated in Eq.~\eqref{eq:critical_condition} and drawn in red. Note that there is no sharp transition here, and the shifts close to the line and in the normal region can still be very large. The blue line going through anomalous regions separates parameters for which the order of the right and the left sideband is inverted and those for which it is not.
    As $|\eta|$ is increased, the cavity is very nonlinear and one may see double resonances and higher-order nonlinearities, among other phenomena. An additional increase leads to the appearance of two solutions, which define the bistability region. In very slow longitudinal fields, the sidebands become invisible due to broadening (the red area).}  
\label{fig:nonlinearity_diagram}
\end{figure}

The conclusions of the section are summarized in Fig.~\ref{fig:nonlinearity_diagram}. The figure sketches different regimes as a function of the cavity nonlinearity strength $\eta$ and the normalized frequency of the longitudinal drive $\omega_{\mathrm{l}}/{\gamma_{\mathrm{c}}}$. The results were generalized to the case of a negative Kerr coefficient ($\eta<0$). The figure is mirror-symmetric with respect to the $\eta$-axis. While we restricted the prior analysis to $\eta>0$, none of the equations rely on that assumption. The difference is that when $\eta<0$ the cavity frequency is red-shifted and not blue-shifted. The anomalous behavior here is that the left sideband in Fig.~\ref{fig:two_effect}(a) is to the right of the central resonance. We call this the {\em left-sideband anomaly} to differentiate it from the case of a positive Kerr coefficient where we call the effect the {\em right-sideband anomaly}.

The red and blue lines in Fig.~\ref{fig:nonlinearity_diagram} represent the conditions~\eqref{eq:critical_condition} and \eqref{eq:inversion_condition}, respectively. The lines are drawn as dashed for high $|\eta|$ to indicate that the approximation of small $|\eta|$ used in this section starts to break down. In the region of strong nonlinearity (the blue area), the spectrum would be significantly altered even before the onset of bistability. Not only do the higher-order terms in $\eta$ become important, but the structure of the spectrum is not well-described by simple resonance conditions. One would have to return to the full analysis of dynamical equations, some of which is presented in App.~\ref{app:spin_cavity}. If the cavity is driven to the nonlinear regime by strong driving $g_{\mathrm{P}}$, the higher-order nonlinearities might also become excited and should be included in the Hamiltonian. The region of small $\omega_{\mathrm{l}}$ is drawn in red and easily matches the condition for the anomaly in Eq.~\eqref{eq:critical_condition}. However, this region is not experimentally accessible since for small $\omega_{\mathrm{l}}$ it becomes impossible to distinguish between the central resonance at $\omega_{\mathrm{t}}$ and the sidebands at $\omega_{\mathrm{t}}\pm\omega_{\mathrm{l}}$.

\section{\label{sec:discussion}Outlook}

Throughout the paper, we treated dissipation in an \emph{ad hoc} way, which is justified in certain limits~\cite{geva1995}. However, more systematic treatment through a Lindblad equation as in Ref.~\cite{greenberg2007} (or alternatively a non-Markovian quantum master equation) might uncover novel details. This relies on a better microscopic understanding of the dissipation in diamond defects. The analysis in the dressed-atom picture should give an intuitive description near the Rabi regime~\cite{greenberg2007}, as well as provide us with a proper extension of the theory with the full quantum treatment of electromagnetic radiation~\cite{cohen1989, carmichael1976}.  

We modeled the cavity nonlinearity as a Kerr nonlinearity due to its relevance for many quantum resonators in solid-state physics. However, we believe that the sideband sensitivity to cavity nonlinearities is more general, and these details remain to be characterized.

\section{Acknowledgments}

We are grateful to Ari Turner for many insightful discussions.  D.P. acknowledges support from the Israel Science Foundation (ISF), grant No.~2005/23.

\section{Author contribution statements}

L.A. conducted the theoretical analysis and wrote the manuscript.  S.H. and S.M. performed the experiments.  D.P. contributed to the theorical work, and  E.B. assisted with the experiment.  

\appendix

\section{Driven spin model}\label{app:spin}

The steady-state solution for the central frequency does not have corrections in linear response,
\begin{equation}\label{eq:occupation_zero}
    \begin{split}
        &\langle\sigma_{\mathrm{z}}\rangle^{(0)} = \frac{\left(\Delta_{\mathrm{s}}^2+\gamma_2^2\right)\sigma_{\mathrm{z}}^{\text{eq}}}{\Delta_{\mathrm{s}}^2+\gamma_2^2+4\frac{\gamma_2}{\gamma_1}|B_{\mathrm{t}}|^2}, \\
        &\langle\sigma_{+}\rangle^{(0)} = \frac{B_{\mathrm{t}}}{\Delta_{\mathrm{s}}+i\gamma_2}\langle\sigma_{\mathrm{z}}\rangle^{(0)}=\frac{\left(\Delta_{\mathrm{s}}-i\gamma_2\right)B_{\mathrm{t}}\sigma_{\mathrm{z}}^{\text{eq}}}{\Delta_{\mathrm{s}}^2+\gamma_2^2+4\frac{\gamma_2}{\gamma_1}|B_{\mathrm{t}}|^2}.
    \end{split}
\end{equation}
After enough time passes, the spin will settle to oscillate at $\pm\omega_{\mathrm{l}}$ frequencies in the rotating frame. We can perform calculations for one of the sideband,s and the other can be obtained by changing $\omega_{\mathrm{l}}$ to $-\omega_{\mathrm{l}}$. Picking up the terms that oscillate at $\omega_{\mathrm{l}}$ we get the following system of equations:
\begin{equation}
    \begin{split}
        &i\frac{d\langle\sigma_+\rangle^{(1)}}{dt}=-\left(\Delta_{\mathrm{s}}-\omega_{\mathrm{l}}+i\gamma_2\right)\langle\sigma_+\rangle^{(1)}\\
        &+B_{\mathrm{t}} \langle\sigma_{\mathrm{z}}\rangle^{(1)}-\frac{B_{\mathrm{l}}}{2}\langle\sigma_+\rangle^{(0)}, \\
        &i\frac{d\langle\sigma_-\rangle^{(1)}}{dt}=\left(\Delta_{\mathrm{s}}+\omega_{\mathrm{l}}-i\gamma_2\right)\langle\sigma_-\rangle^{(1)}\\
        &-B_{\mathrm{t}}^* \langle\sigma_{\mathrm{z}}\rangle^{(1)}+\frac{B_{\mathrm{l}}}{2}\langle\sigma_-\rangle^{(0)}, \\
       &i\frac{d\langle\sigma_{\mathrm{z}}\rangle^{(1)}}{dt}=\left(\omega_{\mathrm{l}}-i\gamma_1\right)\langle\sigma_{\mathrm{z}}\rangle^{(1)}\\
       &+2\left(B_{\mathrm{t}}^*\langle\sigma_+\rangle^{(1)}-B_{\mathrm{t}}\langle\sigma_-\rangle^{(1)}\right).
    \end{split}
\end{equation}
The steady-state solution of the system of equations has two contributions,
\begin{equation}\label{eq:splus1}
    \langle\sigma_+\rangle^{(1)}=\frac{1}{\Delta_{\mathrm{s}}-\omega_{\mathrm{l}}+i\gamma_2}\left(-\frac{B_{\mathrm{l}}}{2}\langle\sigma_+\rangle^{(0)}+B_{\mathrm{t}} \langle\sigma_{\mathrm{z}}\rangle^{(1)}\right).
\end{equation}
The first term in the equation contains two resonances that correspond to two different processes in the linear response regime, where one of the resonances becomes power-broadened (shown in Fig. \ref{fig:evolution}). The other term comes from the fact that the ground and excited state populations start oscillating with $\omega_{\mathrm{l}}$ frequency. The solution is given in the main part of the paper.

\section{Response function poles}\label{app:complex_poles}

\begin{figure*}
    \centering
    \includegraphics[width=.95\textwidth]{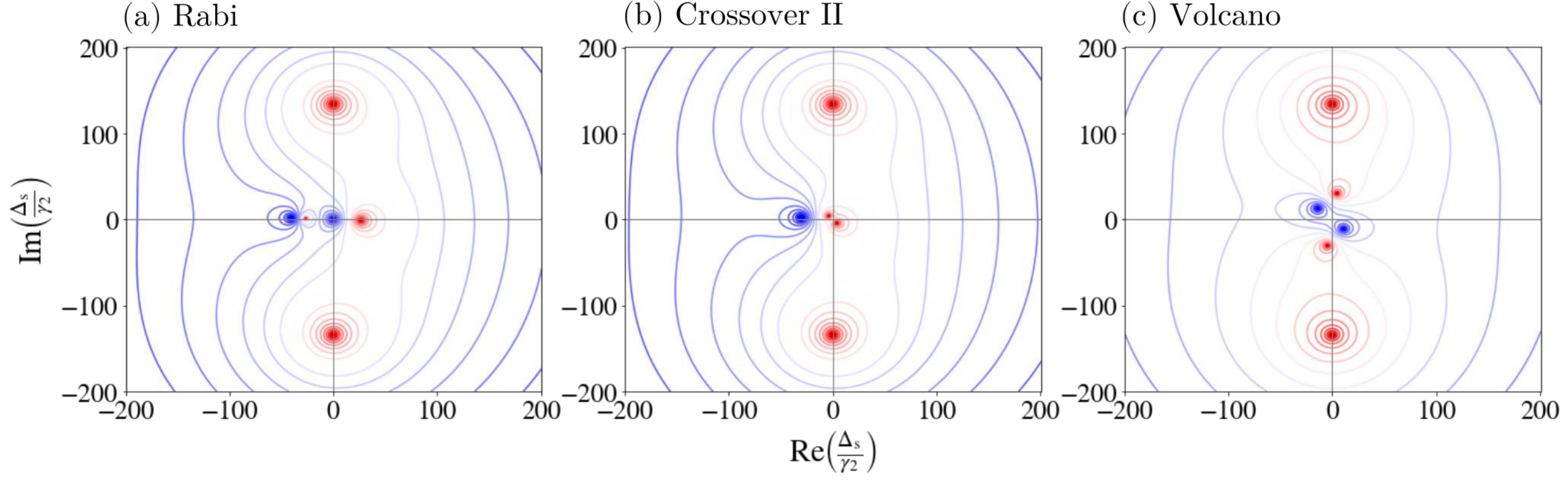}
    \caption{Logarithmic intensity plot of poles associated with resonances (red dots) and zeros (blue dots) as they move through the complex frequency plane. The parameters are: $\gamma_1=0.05$, $|B_{\mathrm{t}}|=15.0$, and $\omega_{\mathrm{l}}$ goes through values $40.0$, $30.0$, $3.0$ from a) to c). Note that in the crossover region in b), where $\omega_{\mathrm{l}}=2B_{\mathrm{t}}$ the two resonances hide the zero at the origin.}
    \label{fig:complex_mov}
\end{figure*}

In this appendix, we present an alternative description of the transition between the Rabi and the volcano regimes. If we make an analytic continuation of $\Delta_{\mathrm{s}}$ to the complex plane, we can track the motion of poles of the response function in the complex plane.  There are four poles in the approximate response, Eq.~(\ref{eq:simplified}), two Rabi poles that are at $\pm\sqrt{\omega_{\mathrm{l}}^2-4|B_{\mathrm{t}}|^2}$, and two poles from the power-broadened part, $\pm 2 i\sqrt{\frac{\gamma_2}{\gamma_1}}|B_{\mathrm{t}}|$, that always lie on the imaginary axis.  The actual positions of the poles are slightly shifted relative to these values, due to the decay rates appearing in the exact expression, Eq.~(\ref{eq:fullequation}).

As shown in Fig.~\ref{fig:complex_mov}, the transition from the Rabi to the volcano regime can be seen as movement of the Rabi resonances in the complex plane away from the real axis as $\omega_{\mathrm{l}}$ is decreased. In the Rabi regime, $2|\Bt|/\oml\ll 1$,  the resonances (red dots near the real axis in Fig.~\ref{fig:complex_mov}(a)) are responsible for most of the weight of the spectrum and they are visible as sharp peaks in the spectrum. They never intersect the real axis, because they always have an imaginary part set by decay rates. As can be seen from Eq.~\eqref{eq:simplified}, the response has zeros at $\Delta_{\mathrm{s}}=0$ and $\Delta_{\mathrm{s}}=-\omega_{\mathrm{l}}$. Since the resonance with the negative real part is close to the zeros it is suppressed relative the other resonance. 

As the ratio of transverse driving to longitudinal frequency $|\Bt|/\oml$ is increased, both resonances move towards the origin, $\Delta_{\mathrm{s}}=0$, leading to the crossover II region [Fig. \ref{fig:complex_mov}(b)]. As $|\Bt|/\oml$ is further increased, the resonances start moving away from each other, mainly along the imaginary axis: now the Rabi condition, \eqref{eq:HHresonance}, is satisfied but for imaginary frequencies. In the volcano regime (shown in \ref{fig:complex_mov}(c)), the imaginary Rabi resonances create a broad peak, and the zeros (blue dots in the figure) create the central transparency region. In addition, the power-broadening resonances that lie further away on the imaginary axis provide weight to the tails of the spectrum and set the overall width of the volcano.

\section{Driven spin-cavity model}\label{app:spin_cavity}

The system of equations for the five operators of the coupled spin-cavity system is:
\begin{equation}\label{eq:full_system}
\begin{split}
&i\frac{d\sigma_+^i}{dt} = -(\Delta_{\mathrm{s}}+B_{\mathrm{l}}\cos{\left(\omega_\mathrm{l} t\right)}+i\gamma_2) \sigma_{+}^i+\Omega a^{\dagger}\sigma_\mathrm{z}^i + \mathcal{F}_+^i, \\
&i\frac{d\sigma_-^i}{dt} = (\Delta_{\mathrm{s}}+B_\mathrm{l}\cos{\left(\omega_\mathrm{l} t\right)}-i\gamma_2) \sigma_{-}^i-\Omega a\sigma_\mathrm{z}^i+\mathcal{F}_-^i, \\
&i\frac{d\sigma_\mathrm{z}^i}{dt}= 2\Omega\left(a\sigma_{+}^i-a^{\dagger}\sigma_{-}^i \right)-i\gamma_{1}\left(\sigma_{\mathrm{z}}^i-\sigma_{\mathrm{z}}^{\mathrm{eq}}\right)+\mathcal{F}_\mathrm{z}^i,\\   
&i\frac{da^{\dagger}}{dt} = -\left(\Delta_{\mathrm{c}}+i\gamma_c\right)a^{\dagger}-K_ca^{\dagger}a^{\dagger}a-\sum_{i=1}^{N_\mathrm{S}}\Omega\sigma_+^i -\frac{g_{\mathrm{P}}}{2}+\mathcal{F}_a^{\dagger}, \\
&i\frac{da}{dt} = \left(\Delta_
{\mathrm{c}}-i\gamma_\mathrm{c}\right)a+K_\mathrm{c}a^{\dagger}aa+\sum_{i=1}^{N_\mathrm{S}}\Omega \sigma_-^i +\frac{g_{\mathrm{P}}}{2}+\mathcal{F}_a,
\end{split}
\end{equation}
where we added the Langevin forces ($\mathcal{F}_i$) to the equations, with the effect that the Lie algebra structure is preserved. The quantum correlations between the cavity and the spin are neglected ($\langle AB\rangle\approx\langle A\rangle\langle B\rangle$). It is clear that every spin within this approximation is equivalent and we can write the equations for an average spin operator components ($\sigma_{\alpha}=\frac{1}{N_\mathrm{S}}\sum_{i=1}^{N_\mathrm{S}}\sigma_{\alpha}^i$). The system of equations for expectation values can be written in terms of dimensionless variables and parameters defined in the following way:
\begin{equation}\label{eq:parameters}
\begin{split}
        \tilde{\Delta}_\mathrm{c}=\frac{\Delta_{\mathrm{c}}}{\gamma_\mathrm{c}},\hspace{2mm} \tilde{\Delta}_{\mathrm{s}}=\frac{\Delta_{\mathrm{s}}}{\gamma_2},\hspace{2mm} \tilde{\gamma}_1=\frac{\gamma_1}{\gamma_2}&, \hspace{2mm} \tilde{\gamma}_\mathrm{c}=\frac{\gamma_\mathrm{c}}{\gamma_2}; \\
        \eta = \frac{g_\mathrm{P}^2K_\mathrm{c}}{4\gamma_\mathrm{c}^3};\hspace{2mm} \nu=\frac{N_\mathrm{S}\Omega^2\sigma_\mathrm{z}^{\mathrm{eq}}}{\gamma_2\gamma_\mathrm{c}}&,\hspace{2mm} \xi = \frac{g_\mathrm{P}^2\Omega^2}{\gamma_1\gamma_2\gamma_\mathrm{c}^2};\\
        A = \frac{\langle a^{\dagger}\rangle}{\frac{g_\mathrm{P}}{2\gamma_\mathrm{c}}},\hspace{2mm} \tilde{\sigma}_\mathrm{z} =\frac{\langle\sigma_\mathrm{z}\rangle}{\sigma_\mathrm{z}^{\mathrm{eq}}}, \hspace{2mm} \tilde{\sigma}_{\pm}&=\frac{\langle\sigma_{\pm}\rangle}{\frac{g_\mathrm{P}\Omega\sigma_\mathrm{z}^{\mathrm{eq}}}{2\gamma_2\gamma_\mathrm{c}}};\\
        \tilde{B}_{\mathrm{l}}=\frac{B_{\mathrm{l}}}{\gamma_2},\ \tilde{\omega}_{\mathrm{l}}=\frac{\omega_{\mathrm{l}}}{\gamma_2}.
\end{split}
\end{equation}
The system of equations with these new variables and parameters is

\begin{eqnarray}\label{eq:system}
    & i\frac{d\tilde{\sigma}_+}{d\tilde{t}} = -\left(\tilde{\Delta}_{\mathrm{s}}+\tilde{B}_\mathrm{l}\cos{\left(\tilde{\omega}_\mathrm{l} \tilde{t}\right)}+i\right)\tilde{\sigma}_++A\tilde{\sigma}_\mathrm{z}, \\
    & i\frac{1}{\tilde{\gamma}_1}\frac{d\tilde{\sigma}_\mathrm{z}}{d\tilde{t}} = \frac{\xi}{2}\left(A^{*}\tilde{\sigma}_+-A\tilde{\sigma}_-\right)-i\left(\tilde{\sigma}_\mathrm{z}-1\right), \\
    & i\frac{1}{\tilde{\gamma}_\mathrm{c}}\frac{d A}{d\tilde{t}} = -\left(\tilde{\Delta}_{\mathrm{c}}+i\right)A-\eta|A|^2A-\nu\tilde{\sigma}_+-1,
\end{eqnarray}
where time is measured in units of $T_2=\frac{1}{\gamma_2}$.

\subsection{Steady-state sideband symmetry \label{app:sideband_symmetry}}
In the absence of cavity nonlinearity ($\eta=0$) there is a symmetry that connects the left and the right sideband. To study the sidebands we are going to write the steady state equations for the $+\omega_\mathrm{l}$ Fourier component,

\begin{equation}
\begin{split}
    &\tilde{\sigma}_+^{(1)}=\frac{-A^{(0)}\tilde{\sigma}_\mathrm{z}^{(1)}-A^{(1)}\tilde{\sigma}_\mathrm{z}^{(0)}+\frac{\tilde{B}_{\mathrm{l}}}{2}\tilde{\sigma}_+^{(0)}}{\left(\tilde{\omega}_{\mathrm{l}}-\tilde{\Delta}_{\mathrm{s}}-i\right)}, \\
    &\tilde{\sigma}_-^{(1)}=\frac{A^{*(0)}\tilde{\sigma}_\mathrm{z}^{(1)}+A^{*(1)}\tilde{\sigma}_\mathrm{z}^{(0)}-\frac{\tilde{B}_{\mathrm{l}}}{2}\tilde{\sigma}_-^{(0)}}{\left(\tilde{\omega}_{\mathrm{l}}+\tilde{\Delta}_{\mathrm{s}}-i\right)}, \\
    &\tilde{\sigma}_\mathrm{z}^{(1)}=-\frac{\frac{\xi}{2}\left(A^{* (0)}\tilde{\sigma}_+^{(1)}-A^{(0)}\tilde{\sigma}_-^{(1)}+A^{* (1)}\tilde{\sigma}_+^{(0)}-A^{(1)}\tilde{\sigma}_-^{(0)}\right)}{\left(\frac{\tilde{\omega}_{\mathrm{l}}}{\tilde{\gamma}_1}-i\right)},\\
    &A^{(1)}=\frac{\nu \tilde{\sigma}_+^{(1)}}{\left(\frac{\tilde{\omega}_{\mathrm{l}}}{\tilde{\gamma}_\mathrm{c}}-\tilde{\Delta}_{\mathrm{c}}-i\right)}, \\
    &A^{*(1)}=-\frac{\nu \tilde{\sigma}_-^{(1)}}{\left(\frac{\tilde{\omega}_{\mathrm{l}}}{\tilde{\gamma}_\mathrm{c}}+\tilde{\Delta}_{\mathrm{c}}-i\right)}.
\end{split}
\end{equation}
Index in brackets specifies the Fourier component of the variables, 0 standing for the central frequency and $\pm 1$ for the right and left sideband frequencies $\pm \omega_{\mathrm{l}}$. If we reverse the signs of the detuning parameters ($\Delta_{\mathrm{s}} \rightarrow -\Delta_{\mathrm{s}}$ and $\Delta_{\mathrm{c}} \rightarrow -\Delta_{\mathrm{c}}$) and change the signs of the variables we can move from one sideband to the other. Transformations

\begin{equation}
    \begin{split}
        &\left(\tilde{\sigma}_i^{(0)}\left(-\Delta_{\mathrm{s}}, -\Delta_{\mathrm{c}}\right)\right)^*=-\tilde{\sigma}_i^{(0)}\left(\Delta_{\mathrm{s}}, \Delta_{\mathrm{c}}\right),\\ &\left(A^{(0)}\left(-\Delta_{\mathrm{s}}, -\Delta_{\mathrm{c}}\right)\right)^*=A^{(0)}\left(\Delta_{\mathrm{s}}, \Delta_{\mathrm{c}}\right),\\
        &\left(A^{*(0)}\left(-\Delta_{\mathrm{s}}, -\Delta_{\mathrm{c}}\right)\right)^*=A^{*(0)}\left(\Delta_{\mathrm{s}}, \Delta_{\mathrm{c}}\right),\\    &\left(\tilde{\sigma}_i^{(-1)}\left(-\Delta_{\mathrm{s}}, -\Delta_{\mathrm{c}}\right)\right)^*=\tilde{\sigma}_i^{(1)}\left(\Delta_{\mathrm{s}}, \Delta_{\mathrm{c}}\right),\\ &\left(A^{(-1)}\left(-\Delta_{\mathrm{s}}, -\Delta_{\mathrm{c}}\right)\right)^*=-A^{(1)}\left(\Delta_{\mathrm{s}}, \Delta_{\mathrm{c}}\right),\\
        &\left(A^{*(-1)}\left(-\Delta_{\mathrm{s}}, -\Delta_{\mathrm{c}}\right)\right)^*=-A^{*(1)}\left(\Delta_{\mathrm{s}}, \Delta_{\mathrm{c}}\right),
    \end{split}
\end{equation}
allow us to find the left sideband from the right sideband. If we are looking at the absolute expectation values of the operators, the transformation reduces to switching the signs of detunings. Turning on the cavity nonlinearity by making $\eta \neq0$ is one possible way to break this symmetry, which leads to a different response for the left and the right sideband.

\subsection{The central resonance}\label{app:central}

We perform the analysis in the dimensionless system of units which elucidates a clear separation between different effects. The frequencies and the rates are expressed in units of the decoherence rate $\gamma_2$ and the cavity width $\gamma_{\mathrm{c}}$,
\begin{equation}\label{eq:parameters}
\begin{split}
        &\tilde{\Delta}_\mathrm{c}=\frac{\Delta_{\mathrm{c}}}{\gamma_\mathrm{c}},\ \tilde{\Delta}_{\mathrm{s}}=\frac{\Delta_{\mathrm{s}}}{\gamma_2},\ \tilde{\gamma}_1=\frac{\gamma_1}{\gamma_2}, \\
        &\tilde{\gamma}_\mathrm{c}=\frac{\gamma_\mathrm{c}}{\gamma_2}, \ 
        \tilde{B}_{\mathrm{l}}=\frac{B_{\mathrm{l}}}{\gamma_2},\ \tilde{\omega}_{\mathrm{l}}=\frac{\omega_{\mathrm{l}}}{\gamma_2},\ \tilde{t}=\gamma_2 t.
\end{split}
\end{equation}
The rescaled variables $\tilde{\sigma}_i$ and $A$ correspond to the spin and the cavity, respectively, where
\begin{equation}
 A = \frac{\langle a^{\dagger}\rangle}{\frac{g_\mathrm{P}}{2\gamma_\mathrm{c}}},\ \tilde{\sigma}_\mathrm{z} =\frac{\langle\sigma_\mathrm{z}\rangle}{\sigma_\mathrm{z}^{\mathrm{eq}}}, \ \tilde{\sigma}_{\pm}=\frac{\langle\sigma_{\pm}\rangle}{\frac{g_\mathrm{P}\Omega\sigma_\mathrm{z}^{\mathrm{eq}}}{2\gamma_2\gamma_\mathrm{c}}}.
\end{equation}
The system of equations with new parameters is: 
\begin{subequations}
\label{eq:dimensionless_system} 
\begin{align}\label{eq:dimensionless_system1}
    & i\frac{d\tilde{\sigma}_+}{d\tilde{t}} = -\left(\tilde{\Delta}_{\mathrm{s}}+\tilde{B}_\mathrm{l}\cos{\left(\tilde{\omega}_\mathrm{l} \tilde{t}\right)}+i\right)\tilde{\sigma}_++A\tilde{\sigma}_\mathrm{z}, \\
    &\label{eq:dimensionless_system2} i\frac{1}{\tilde{\gamma}_1}\frac{d\tilde{\sigma}_\mathrm{z}}{d\tilde{t}} = \frac{\xi}{2}\left(A^{*}\tilde{\sigma}_+-A\tilde{\sigma}_-\right)-i\left(\tilde{\sigma}_\mathrm{z}-1\right), \\
    &\label{eq:dimensionless_system3} i\frac{1}{\tilde{\gamma}_\mathrm{c}}\frac{d A}{d\tilde{t}} = -\left(\tilde{\Delta}_{\mathrm{c}}+i\right)A-\eta|A|^2A-\nu\tilde{\sigma}_+-1.
\end{align}
\end{subequations}
The dimensionless parameters $\eta$, $\nu$, and $\xi$ set the strengths of different effects,
\begin{equation}
    \eta = \frac{g_\mathrm{P}^2K_\mathrm{c}}{4\gamma_\mathrm{c}^3},\ \nu=\frac{N_\mathrm{S}\Omega^2\sigma_\mathrm{z}^{\mathrm{eq}}}{\gamma_2\gamma_\mathrm{c}},\ \xi = \frac{g_\mathrm{P}^2\Omega^2}{\gamma_1\gamma_2\gamma_\mathrm{c}^2}.
\end{equation}
The parameter $\eta$ determines the amplitude of \emph{the cavity (Kerr) nonlinearity}, while $\nu$ determines the strength of the \emph{the spin-cavity feedback}. Parameter $\xi$ determines the scale of the power broadening.

We first review the effect of the nonlinearities on the central resonance of the cavity at angular frequency $\omega_{\mathrm{t}}$ and proceed to discuss the results for the nonlinear response at sideband frequencies $\omega_{\mathrm{t}}\pm \omega_{\mathrm{l}}$.

The central resonance is obtained by looking at the zero-frequency component of equations~\eqref{eq:dimensionless_system}. When a linear cavity ($\eta=0$) is uncoupled from the spin ($\nu=0$) it settles to a steady state where the number of photons in the cavity is determined by how strongly it is driven (set by $g_{\mathrm{P}}$ and $\Delta_{\mathrm{c}}$) and how fast it is losing photons ($\gamma_{\mathrm{c}}$). The nonlinearities have the effect of changing the shape of the spectral line and renormalizing the resonant frequency of the cavity. The central-resonance steady state of the cavity is the solution of the polynomial equation,
\begin{equation}\label{eq:central_cavity_equation}
    A^{(0)} = \frac{-1}{\tilde{\Delta}_{\mathrm{c}}+i+\eta |A^{(0)}|^2+\frac{\nu\left({\tilde{\Delta}_{\mathrm{s}}}-i\right)}{1+\tilde{\Delta}_{\mathrm{s}}^2+\xi|A^{(0)}|^2}}.
\end{equation}

\begin{figure}
    \centering
    \includegraphics[width=0.49\textwidth]{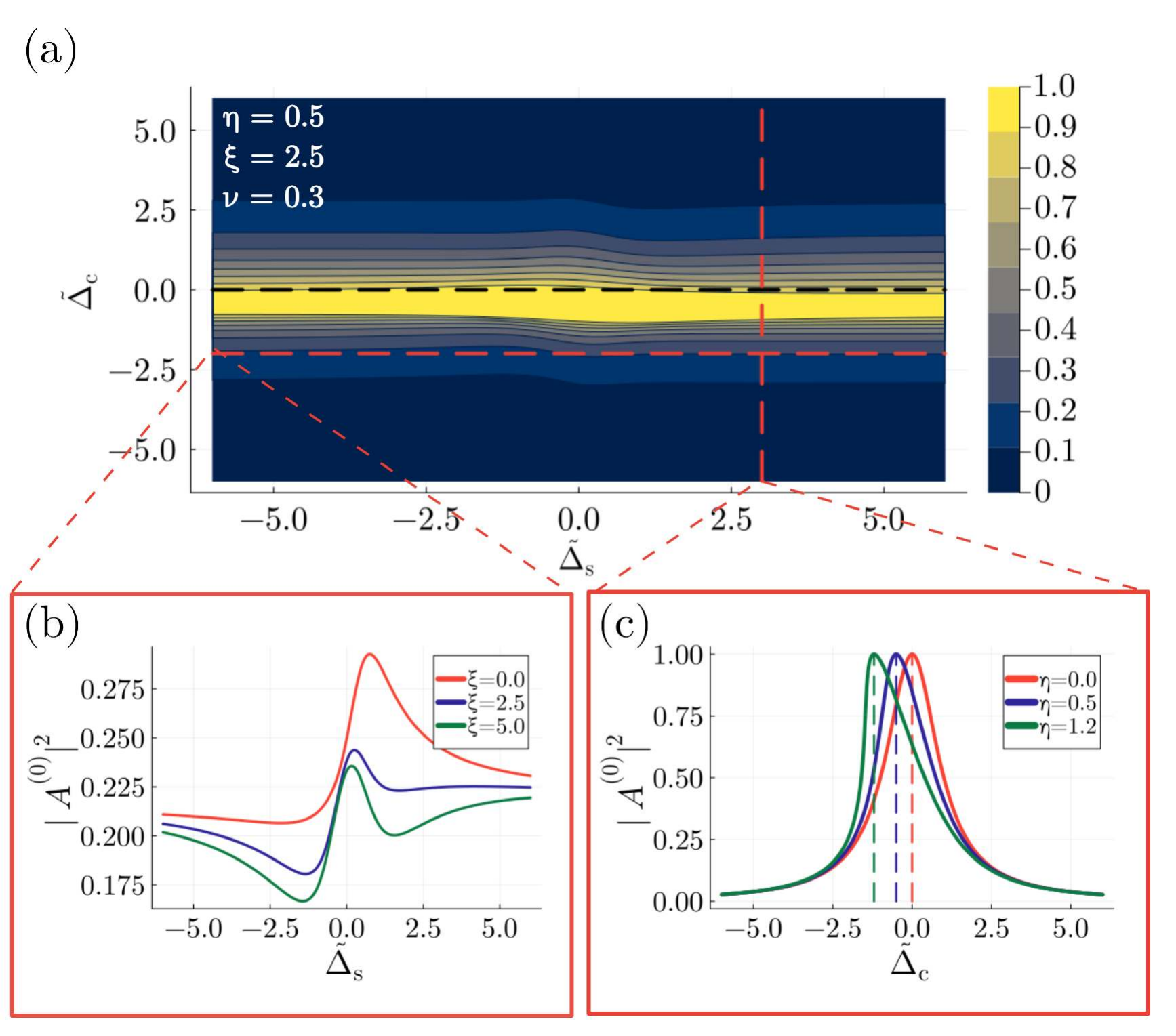}
    \caption{The effect of both nonlinearities on the central resonance is presented in panel (a) as a two-dimensional plot in the $\tilde{\Delta}_{\mathrm{s}}-\tilde{\Delta}_{\mathrm{c}}$ plane for parameters shown in the plot. The $S$-shaped spectral line as a function of spin detuning $\tilde{\Delta}_\mathrm{s}$ is shown in panel (b) at $\tilde{\Delta}_{\mathrm{c}}=-2.0$ for different values of $\xi$. In (c), the spin-cavity feedback is ignored ($\nu=0$), and $\tilde{\Delta}_{\mathrm{s}}=3.0$. The maximum of the resonance is shown to closely follow an approximate result $-\eta$ for different values of $\eta$.}  
\label{fig:combined_center}
\end{figure}

If both the spin and the Kerr nonlinearity are combined, we need to solve for the roots of the quintic equation, obtained by taking a modulus of Eq.~\eqref{eq:central_cavity_equation}. The numerical solution is displayed in Fig. ~\ref{fig:combined_center}(a) where the central resonance intensity is plotted in the $\tilde{\Delta}_{\mathrm{s}}-\tilde{\Delta}_{\mathrm{c}}$ plane. 

The spin-cavity feedback changes spectral properties of the cavity near the spin resonance with the appearance of the characteristic $S$-shape in the dependence of the spectral intensity on the spin detuning $\tilde{\Delta}_{\mathrm{s}}$. The parameter $\nu$ determining the strength of this effect is proportional to the number of spins $N_{\mathrm{S}}$ interacting with the cavity. If a lot of spins interact with the cavity, they visibly alter the spectrum near $\tilde{\Delta}_{\mathrm{s}}=0$. The dependence of the $S$-shaped response on the parameter $\xi$ is shown in Fig.~\ref{fig:combined_center}(b). As $\xi$ is increased, due to the increasing power of the drive, the spins effectively decouple from the cavity, since the sharp resonance near $\tilde{\Delta}_{\mathrm{s}}$ is washed out.

Without nonlinearities, the cavity is a simple Lorentzian,
\begin{equation}
    |A^{(0)}|^2 = \frac{1}{1+\tilde{\Delta}_{\mathrm{c}}^2}\Rightarrow B_{\mathrm{t}}=\Omega|\langle a^{\dagger (0)}\rangle|=\frac{g_\mathrm{P}^2}{4\left(\Delta_{\mathrm{c}}^2+\gamma_{\mathrm{c}}^2\right)},
\end{equation}
where $\Omega|\langle a^{\dagger (0)}\rangle|$ acts as an effective magnetic field, as shown by the system of equations in~\eqref{eq:full_system}. Nonlinearities may induce instabilities, which appear in Eq.~\eqref{eq:central_cavity_equation} when the quintic polynomial for $|A^{(0)}|^2$ acquires multiple positive roots. Although we study the cavity beyond the linear regime, we will not address the instability region. 

Since the spin nonlinearity is localized near the spin resonance ($\tilde{\Delta}_{\mathrm{s}}=0$), far away from this resonance ($\tilde{\Delta}_{\mathrm{s}}\gg 1$) there is only a Kerr nonlinearity and the central frequency solution is approximately given as
\begin{equation}
    |A^{(0)}|^2\rightarrow\frac{1}{\left(\tilde{\Delta}_{\mathrm{c}}+\eta|A^{(0)}|^2\right)^2+1}, \hspace{2mm}\mathrm{when}\; \tilde{\Delta}_{\mathrm{s}}\rightarrow \infty.
\end{equation}
The physically relevant solution is the real root of this cubic equation and there is only one real root for any $\tilde{\Delta}_{\mathrm{c}}$ below $\eta \approx 1.54$. The position of the peak is the solution of:
\begin{equation}
    \tilde{\Delta}_{\mathrm{c}}+\eta|A^{(0)}|^2=0,
\end{equation}
To cubic order in $\eta$ the position of the center is shifted to

\begin{equation}
    \tilde{\Delta}_{\mathrm{c,\;res}}^{(0)}|_{\tilde{\Delta}_{\mathrm{s}}\rightarrow\infty}=-\eta+\mathcal{O}\left(\eta^3\right),
\end{equation}

On the other hand, without the Kerr nonlinearity ($\eta=0$), the central frequency response is a solution to another cubic equation, 
\begin{equation}
    |A^{(0)}|^2=\left|\frac{-1}{\tilde{\Delta}_{\mathrm{c}}+i+\frac{\nu\left(\tilde{\Delta}_{\mathrm{s}}-i\right)}{1+\tilde{\Delta}_{\mathrm{s}}^2+\xi|A^{(0)}|^2}}\right|^2, \hspace{2mm} \mathrm{as}\;\eta \rightarrow 0.
\end{equation}
When the spin’s feedback into the cavity is weak, the solution linear in $\nu$ remains a good approximation,
\begin{equation}
    |A^{(0)}|^2\approx\frac{1}{1+\tilde{\Delta}_{\mathrm{c}}^2}\left(1-\frac{2\nu\left(\tilde{\Delta}_{\mathrm{c}}\tilde{\Delta}_{\mathrm{s}}-1\right)}{\left(1+\tilde{\Delta}_{\mathrm{c}}^2\right)\left(1+\tilde{\Delta}_{\mathrm{s}}^2\right)+\xi}\right)+\mathcal{O}\left(\nu^2\right).
\end{equation}
If both of these are combined, we need to solve for the roots of the quintic equation. The results are presented in Fig. \ref{fig:combined_center}.

\subsection{Sidebands}\label{app:sidebands}

Assuming $\nu$ and $\eta$ are small, we start the analysis of the sideband response, including the cavity nonlinearity. We show that the sideband spectral lines exhibit large sensitivity to the nonlinear effects, where a dramatic change of the spectrum results even from a moderately strong nonlinearity. Arguably, the most peculiar effect, which we call \emph{the sideband anomaly}, is the possibility that both of the sidebands emerge on the same side of the central resonance.

We look at the cavity steady-state expectation value component that oscillates at the right sideband frequency ($\omega_{\mathrm{t}}+\omega_{\mathrm{l}}$). To the first order in the longitudinal field amplitude, the steady state, obtained from Eq.~\eqref{eq:dimensionless_system3}, is given by
\begin{equation}\label{eq:A1}
A^{(1)}=\frac{\nu\tilde{\sigma}_+^{(1)}+2\eta|A^{(0)}|^2A^{(1)}+\eta\left(A^{(0)}\right)^2 A^{*(1)}}{\left(\frac{\tilde{\omega}_\mathrm{l}}{\tilde{\gamma}_\mathrm{c}}-\tilde{\Delta}_{\mathrm{c}}-i\right)},
\end{equation}
where the absolute value $|A^{(1)}|$ determines the cavity amplitude at the right sideband frequency $\omega_{\mathrm{t}}+\omega_{\mathrm{l}}$. 

The expression in Eq.~\eqref{eq:A1} encapsulates many effects. Before we turn to the analysis of the second term in the numerator, \emph{i.e.} $2\eta|A^{(0)}|^2A^{(1)}$, we discuss the meaning of other terms.
We assume that the spin-cavity feedback is weak ($\nu\ll 1$). 

The term in the denominator is a resonance that occurs when the sideband frequency hits the cavity resonance ($\Delta_{\mathrm{c}}=\omega_{\mathrm{l}}$). The first term in the numerator, $\nu \tilde{\sigma}_+^{(1)}$, represents the spin that drives the cavity at the right-sideband frequency. The expectation value of the spin operator $\tilde{\sigma}_{+}^{(1)}$ at the right sideband frequency is
\begin{equation}\label{eq:sigmap1_nonlinear}
\tilde{\sigma}_{+}^{(1)}=\frac{\tilde{B}_{\mathrm{l}}\tilde{\sigma}_+^{(0)}-A^{(0)}\tilde{\sigma}_{\mathrm{z}}^{(1)}-A^{(1)}\tilde{\sigma}_{\mathrm{z}}^{(0)}}{\tilde{\omega}_{\mathrm{l}}-\tilde{\Delta}_{\mathrm{s}}-i}.
\end{equation}
One of the effects of the Kerr nonlinearity is felt in the fact that the central resonance $A^{(0)}$ is modified by the Kerr nonlinearity, as shown in Appendix~\ref{app:spin_cavity}. Furthermore, the term $-A^{(1)}\tilde{\sigma}_z^{(0)}$ indicates that the cavity now drives the spin at the sideband frequency. However, when the spin-cavity feedback is negligible this term can be ignored. 

A more impactful consequence of the Kerr nonlinearity comes from the last term in the numerator of Eq.~\eqref{eq:A1}. It connects the right sideband amplitude $A^{(1)}$ equation with its conjugate. In other words, it ``mixes" the right and the left sideband solution. This term does not affect the positions of the sideband resonances, but only their detailed shape.

\subsection{The sideband anomaly}

To find the steady-state solutions for the sidebands, we need to pick up all the terms in the system of equations in Eq.~\eqref{eq:full_system} that oscillate at frequency $\omega_{\mathrm{l}}$. The cavity equations for the right sideband are:
\begin{equation}\label{eq:factortwo}
\begin{split}
    &\left(\frac{\tilde{\omega}_\mathrm{l}}{\tilde{\gamma}_\mathrm{c}}-\tilde{\Delta}_{\mathrm{c}}-i\right)A^{(1)}-2\eta|A^{(0)}|^2A^{(1)}\\
    &-\eta\left(A^{(0)}\right)^2 A^{*(1)}-\nu\tilde{\sigma}_+^{(1)}=0, \\
    &\left(\frac{\tilde{\omega}_\mathrm{l}}{\tilde{\gamma}_\mathrm{c}}+\tilde{\Delta}_{\mathrm{c}}-i\right)A^{*(1)}+2\eta|A^{(0)}|^2A^{*(1)}\\
    &+\eta\left(A^{*(0)}\right)^2 A^{(1)}+\nu\tilde{\sigma}_-^{(1)}=0.
\end{split}
\end{equation}
The previous two equations can be combined into the expression

\begin{equation}
    \begin{split}
    &A^{(1)}=-\frac{\tilde{\Delta}_{\mathrm{c}}-z_-}{\left(\tilde{\Delta}_{\mathrm{c}}-z_+\right)\left(\tilde{\Delta}_{\mathrm{c}}-z_-\right)-\eta^2|A^{(0)}|^4}\\
    &\times\left(\nu\tilde{\sigma}_+^{(1)}-\frac{\eta \left(A^{(0)}\right)^2}{\tilde{\Delta}_{\mathrm{c}}-z_-}\nu\tilde{\sigma}_-^{(1)}\right),
    \end{split}
\end{equation}
where we introduced the notation for the complex frequencies,

\begin{equation}
    z_{\pm}=\pm\frac{\tilde{\omega}_\mathrm{l}}{\tilde{\gamma}_\mathrm{c}}-2\eta|A^{(0)}|^2\mp i.
\end{equation}
The resonance that multiplies the rest of the expression that we denote by $\mathcal{R}$ reduces to
\begin{equation}
    \mathcal{R}\underset{\eta\rightarrow 0}{\rightarrow} \frac{1}{\tilde{\Delta}_{\mathrm{c}}\mp\frac{\tilde{\omega}_\mathrm{l}}{\tilde{\gamma}_\mathrm{c}}\pm i}=\frac{\gamma_\mathrm{c}}{\Delta_{\mathrm{c}}\mp\omega_\mathrm{l}\pm i\gamma_\mathrm{c}},
\end{equation}
in the limit $\eta\rightarrow0$. The expression describes an amplification of the sideband radiation when the outgoing light hits the cavity resonantly.

As $\eta$ is increased, the existing resonance is shifted and a new resonance appears. Since we are working inside the stability region, the expression is evaluated for small $\eta$. $\mathcal{R}$ can be written as
\begin{equation}
\begin{split}
    &\mathcal{R}=-\frac{\tilde{\Delta}_{\mathrm{c}}-z_-}{\left(\tilde{\Delta}_{\mathrm{c}}-\tilde{\Delta}_{\mathrm{c},\;+}\right)\left(\tilde{\Delta}_{\mathrm{c}}-\tilde{\Delta}_{\mathrm{c},\;-}\right)}, \\
&\tilde{\Delta}_{\mathrm{c},\;\pm}=\pm\left(\frac{\tilde{\omega}_\mathrm{l}}{\tilde{\gamma}_\mathrm{c}}-i\right)\sqrt{1+\frac{\eta^2 |A^{(0)}|^4}{\left(\frac{\tilde{\omega}_\mathrm{l}}{\tilde{\gamma}_\mathrm{c}}-i\right)^2}}-2\eta|A^{(0)}|^2,
\end{split}
\end{equation}
and for small $\eta$, one of the resonances is dominant,
\begin{equation}
    \mathcal{R}\approx -\left(1-\frac{\frac{\eta^2|A^{(0)}|^4}{2\left(\frac{\tilde{\omega}_\mathrm{l}}{\tilde{\gamma}_\mathrm{c}}-i\right)}}{\tilde{\Delta}_{\mathrm{c}}-z_-+\frac{\eta^2|A^{(0)}|^4}{2\left(\frac{\tilde{\omega}_\mathrm{l}}{\tilde{\gamma}_\mathrm{c}}-i\right)}}\right)\frac{1}{\tilde{\Delta}_{\mathrm{c}}-z_+-\frac{\eta^2|A^{(0)}|^4}{2\left(\frac{\tilde{\omega}_\mathrm{l}}{\tilde{\gamma}_\mathrm{c}}-i\right)}}.
\end{equation}
The center of the resonance is the solution of the equation,

\begin{equation}
    \tilde{\Delta}_{\mathrm{c}}-\frac{\tilde{\omega}_\mathrm{l}}{\tilde{\gamma}_\mathrm{c}}-2\eta|A^{(0)}|^2-\frac{\eta^2|A^{(0)}|^4\frac{\tilde{\omega}_\mathrm{l}}{\tilde{\gamma}_\mathrm{c}}}{2\left(1+\left(\frac{\tilde{\omega}_\mathrm{l}}{\tilde{\gamma}_\mathrm{c}}\right)^2\right)}=0.
\end{equation}
Since $A^{(0)}$ is a function of both $\tilde{\Delta}_{\mathrm{c}}$ and $\eta$, this is a polynomial equation whose real root is the center of the resonance. By performing a similar analysis for the left-sideband solution, we write the general result as
\begin{equation}
    \tilde{\Delta}_{\mathrm{c,\;res}}^{(\pm 1)}=\pm\frac{\tilde{\omega}_\mathrm{l}}{\tilde{\gamma}_\mathrm{c}}-2\eta\frac{1}{1+\left(\frac{\tilde{\omega}_\mathrm{l}}{\tilde{\gamma}_\mathrm{c}}\right)^2}\mp\frac{7}{2}\eta^2\frac{\frac{\tilde{\omega}_\mathrm{l}}{\tilde{\gamma}_\mathrm{c}}}{\left(1+\left(\frac{\tilde{\omega}_\mathrm{l}}{\tilde{\gamma}_\mathrm{c}}\right)^2\right)^3}+\mathcal{O}\left(\eta^3\right).
\end{equation}

\section{The experiment with \label{app:p1nv}P1 and NV centers in diamond}

\begin{figure}
    \centering
    \includegraphics[width=.45\textwidth]{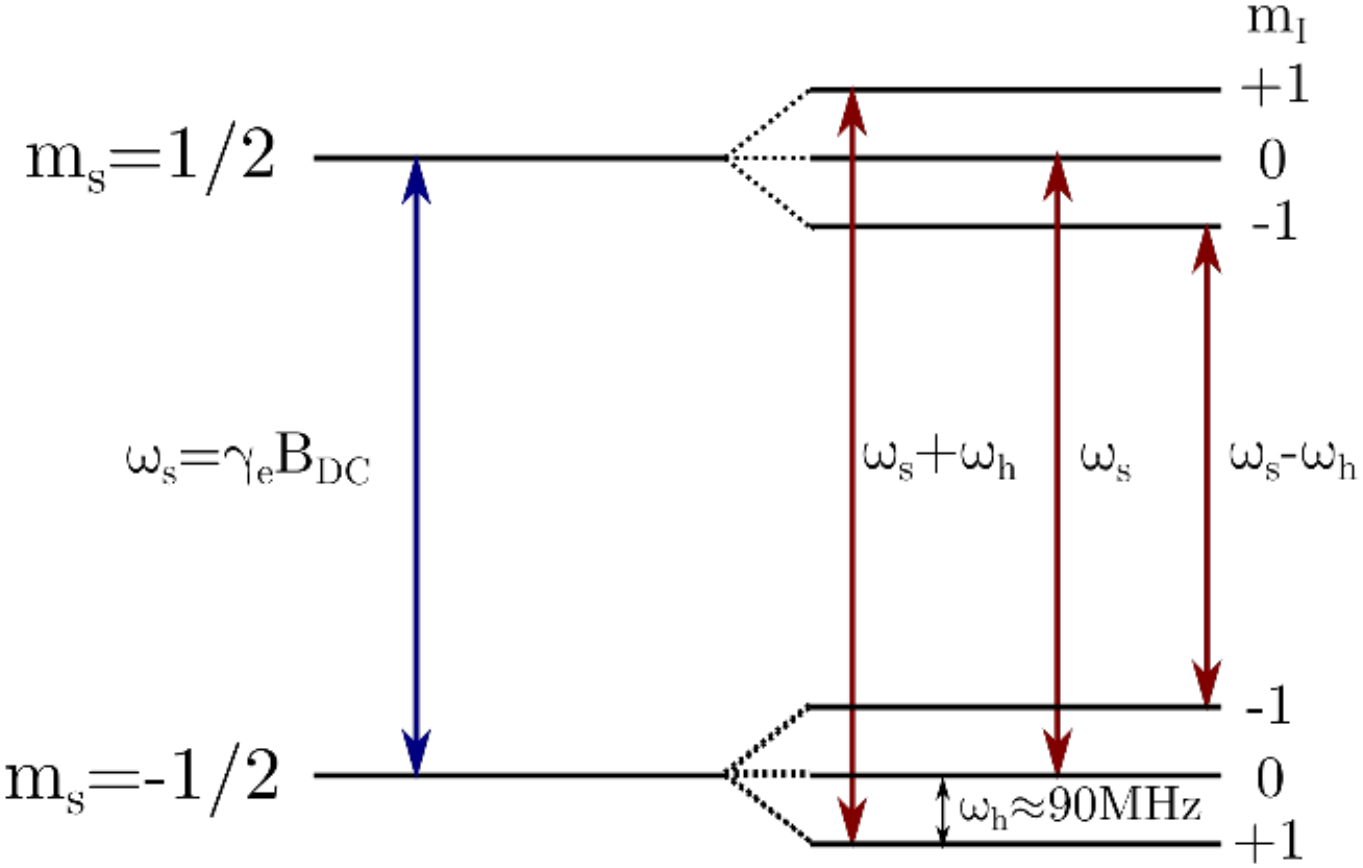}
    \caption{Diagram of the hyperfine structure of the P$1$ center in diamond. The external static field $B_{\mathrm{DC}}$ sets the Zeeman splitting $\omega_{\mathrm{s}}$. The interaction with the nucleus introduces additional structure to the energy-level diagram with the hyperfine splitting $\omega_{\mathrm{h}}$. $m_{\mathrm{s}}$ is the spin quantum number and $m_{\mathrm{I}}$ is the nuclear spin number. The allowed transitions are marked with the red arrows in the diagram.}
    \label{fig:p1_hyperfine}
\end{figure}

We present experimental measurements of the sideband-emission spectrum of the  P$1$ center. We first discuss the discovery of the volcano transparency flanked by asymmetric peaks as seen in Fig.~\ref{fig:experimental_volcano}. We also show that the cavity-based experiment displays a left sideband anomaly both in the P1 and in the NV center.

\subsection{Substitutional nitrogen (P1) defect}

In P$1$ or the substitutional nitrogen defect, a carbon atom is substituted by nitrogen. The unpaired electron forms the defect with one of the four possible neighboring carbon atoms. The spin of the defect lies within a relatively large diamond band gap~\cite{annamafe2018} and interacts with the nitrogen spin-$1$ nucleus which leads to a hyperfine splitting.  The effective Hamiltonian for the isolated spin is
\begin{equation}
    \begin{split}
    H_{\mathrm{P}1} &= \gamma_{\mathrm{e}} B_{\mathrm{DC}} \left(S_{z}\cos{\theta}+S_{x}\sin{\theta}\right)\\
    &+A_{\parallel} S_{z}I_{z}+A_{\perp}\left(S_{x}I_{x}+S_{y}I_{y}\right),
    \end{split}
\end{equation}
where $A_{\parallel}$ and $A_{\perp}$ are parameters that give the strength of the hyperfine coupling parallel and perpendicular to the hyperfine axis, respectively~\cite{simanovskaia2013}. Here, $S_i$ and $I_i$ are spin-1/2 and spin-1 operators, respectively, the $z$-axis is aligned with the defect, and $\theta$ is the angle between the DC field and the defect axis.  If the DC magnetic field is applied along one of the cubic axes of the crystal, then it makes an equal angle $\theta_0=\arccos{\left({1/\sqrt{3}}\right)}$ with respect to all four possible orientations of the defect [see Fig.~\ref{fig:diamond_balls}(b)].  In that case, the response is uniform for all defects.

The DC magnetic field sets the scale of the energy splitting between the hyperfine triplets, $\omega_{\mathrm{s}}=\gamma_\mathrm{e} B_{\mathrm{DC}}$. The strong high-frequency magnetic field acts perpendicular to the DC field and the low-frequency field acts parallel to it. We move to the frame rotating with the transverse frequency $\omega_{\mathrm{t}}$ about the axis of ${\mathbf B}_{\mathrm{DC}}$. The Hamiltonian in the rotating-wave approximation and large $\omega_{\mathrm{s}}$ compared to the hyperfine terms is
\begin{equation}
    \tilde{H}_{\mathrm{P}1} = \left(\Delta_{\mathrm{s}}+B_\mathrm{l}\cos{\left(\omega_\mathrm{l} t\right)}\right) S_{z^\prime} + B_{\mathrm{t}} S_{x^\prime}+\omega_{\mathrm{h}} S_{z^\prime} I_{z^\prime},
\end{equation}
where the primed axes are aligned with the DC field, $\Delta_{\mathrm{s}}=\omega_{\mathrm{s}}-\omega_{\mathrm{t}}$ is the detuning of the spin from the transverse drive. The frequency $\omega_{\mathrm{h}} = \sqrt{A_{\parallel}^2\cos^2{\theta}+A_{\perp}^2\sin^2{\theta}}$ sets the scale of the hyperfine splitting within a triplet. For the specific symmetric angle $\theta_0$ this frequency is the same for all four possible directions, $\omega_{\mathrm{h}}=\sqrt{A_\parallel^2/3+\left(2A_{\perp}^2\right)/3}$. The three transitions allowed by selection rules have frequencies $\omega_{\mathrm{s}}$, and $\omega_{\mathrm{s}}\pm\omega_{\mathrm{h}}$, as shown in Fig.~\ref{fig:p1_hyperfine}.

The defect Hamiltonian effectively splits into three driven non-interacting spin subspaces. The six-dimensional Hilbert space is written as the block-diagonal Hamiltonian,
\begin{equation}
    \tilde{H}_{\mathrm{P}1} = \bigoplus_{n\in \{0, \pm 1\}}H_{\mathrm{s}}^{\mathrm{eff}}(\Delta_{\mathrm{s}}+n\omega_{\mathrm{h}}).
\end{equation}
Each block is the Hamiltonian $H_{\mathrm{s}}^{\mathrm{eff}}$ from Eq.~\eqref{eq:spin_equation} with spin energies $\omega_{\mathrm{s}}$, $\omega_{\mathrm{s}}+\omega_{\mathrm{h}}$, and $\omega_{\mathrm{s}}-\omega_{\mathrm{h}}$. Although these three transitions are independent, the transverse driving field can excite them simultaneously. The intensity of the emitted light is the square of the sum of electromagnetic fields emitted from the defect. Since this field is proportional to the $\sigma_{+}$ operator, we write
\begin{equation}\label{eq:hyperfine_interference}
    \mathcal{I}\propto\left|\sum_{n\in \{0,\pm 1\}}\braket{\sigma_{+}\left(\Delta_{\mathrm{s}}+n\omega_{\mathrm{h}}\right)}\right|^2.
\end{equation}
This allows for the interference between fields that arise from different hyperfine transitions.

Experimentally, we find three volcano-like lineshapes as predicted by theory [see Fig.~\ref{fig:experimental_volcano}]. However, one of the volcanoes in each sideband has the opposite asymmetry.  We argue that this occurs due to the interference effect described by Eq.~\eqref{eq:hyperfine_interference}. We consider a phenomenological model where the longitudinal field causes the target equilibrium polarization in Eq.~(\ref{eq:spin_system}) to be time-dependent,
\begin{equation}\label{eq:time_dependent_equilibrium}
\sigma_{\mathrm{z}}^{\mathrm{eq}}\left(t\right)=\frac{\mathrm{Tr}\left(\sigma_{\mathrm{z}} e^{-\beta \tilde{H}_{\mathrm{P}1}(t)}\right)}{\mathrm{Tr}\left(e^{-\beta \tilde{H}_{\mathrm{P}1}(t)}\right)},
\end{equation}
where $\beta=1/k_B T$ is an effective inverse temperature. The details of the model are discussed in Appendix~\ref{app:thermal}.

For the data in Fig. \ref{fig:experimental_volcano}, the optical laser discussed in Sec.~\ref{sec:exp_setup} is used for cooling, the microwave input power is $-40 \operatorname{dBm}$, and the RF peak-to-peak amplitude is 10 V ($24 \operatorname{dBm}$). The electron-spin resonance at frequency $\omega_\mathrm{s}/(2\pi)$ is split due to interaction with the nuclear spin into three resonances. The splitting extracted from the measurement is $\omega_\mathrm{h}/(2\pi)\approx 90\operatorname{MHz}$. The splitting is significantly better resolved in the sidebands compared to the main peak data [see the top row of Fig.~\ref{fig:experimental_volcano}].

As we see in the experimental data, both sidebands lie to the right of the central resonance frequency $\omega_{\mathrm{c}}/(2\pi)\approx 3.772$ GHz. This corresponds to a right-sideband anomaly.   

As we discussed in Sec.~\ref{subsec:SidebandSymmetry}, this is prohibited in the simple linear spin-cavity model. Therefore, we proposed introducing nonlinearity to break the symmetry between the right and the left sideband. In Sec.~\ref{sec:nonlinear_cavity} we demonstrated a mechanism by which Kerr nonlinearity can lead to this behavior. However, the experiments were performed outside of the region where the cavity nonlinearity is significantly excited, and this anomalous behavior remains to be explained. We now present the sideband data for the NV center that further explores the anomaly.

\subsection{Nitrogen-vacancy (NV$^-$) defect}

In an NV$^-$ defect, the nitrogen that is in place of carbon couples to the vacancy at one of the four possible neighboring positions [see Fig.~\ref{fig:diamond_balls} (a)]. Nitrogen provides two electrons and three carbons contribute each with one electron. The sixth electron that makes the defect negatively charged comes from the environment (possibly from the substitutional nitrogen)~\cite{loubser1978, doherty2013}. The ground state is a spin triplet since the two holes at the defect prefer to form a spatially antisymmetric state to lower their Coulomb energy. The triplet is further split by the dipole-dipole spin interaction, where the spin-$0$ projection is lower in energy [for details see~\cite{maze2011}]. The splitting of the two other levels can be adjusted by the DC magnetic field.

When the defect is driven by two crossed fields and the transverse field is tuned close to the resonance between states with spin projection $S_{z}=\pm 1$, then the $S_z=0$ state can be neglected, and we can effectively treat the defect as a single doubly-driven spin-$1/2$ system. 

The measurements of the NV center are presented in Fig.~\ref{fig:nv_rf_frequency}. We show the left and the right sideband spectra for different longitudinal frequencies $\omega_{\mathrm{l}}$. The optical laser for OISP is on and the microwave input power is $-60\operatorname{dBm}$. The measurements are performed for various longitudinal peak-to-peak amplitudes ($V_{\mathrm{RF}}$ in the figure) and transverse driving frequencies $\omega_{\mathrm{t}}$. All of the sideband resonances lie on the right side of the main peak, which means that they exhibit a right-sideband anomaly, as was found for the P1 centers. As $\omega_{\mathrm{l}}$ is increased the distance between the right and the left sideband increases. We obtained the same result in the theoretical calculation with Kerr nonlinearity in Sec.~\ref{sec:nonlinear_cavity}. 

As the longitudinal frequency is increased, the sideband signal becomes weaker and harder to discern from noise. As the RF amplitude increases (which corresponds to increasing $B_{\mathrm{l}}$) the signal becomes stronger.  In the experiment, the RF field remains in the linear regime such that higher-order sidebands are not observed. 

\subsection{Volcano transparency in P1 centers}\label{app:thermal}

The volcano asymmetry for one of the hyperfine
transitions in the experimental data in Fig.~\ref{fig:experimental_volcano} is opposite to what is expected from theory. Therefore, we propose a modification to the Bloch equations [see Eq.~\eqref{eq:spin_system}] in which we treat the equilibrium polarization as a time-dependent quantity. Since the Hamiltonian oscillates with a slow longitudinal frequency $\omega_{\mathrm{l}}$, we consider relaxation with a constant rate towards a time-dependent polarization, $\sigma_{z}^{\mathrm{eq}}(t)$, given by the instantaneous thermal equilibrium of the Hamiltonian at time $t$, see Eq.~\eqref{eq:time_dependent_equilibrium}. 

We do not provide the microscopic basis for the phenomenological model of Eq.~\eqref{eq:time_dependent_equilibrium}. This is in part because we do not understand well the relaxation processes in diamond defects with the OISP technique even in the absence of the longitudinal field. The model that we use cannot be derived microscopically in the form of the Lindblad equation, which uses the secular approximation. It is desirable to derive the Bloch equations from the master equation in Linbdlad form since the evolution is guaranteed to be trace-preserving and completely positive. The Lindblad equation generalizations beyond the secular approximation (which excludes the oscillating terms) were proposed recently~\cite{mccauley2020, trushechkin2021, potts2021, pradilla2024}, as well as generalizations for driven~\cite{dann2018}, and Floquet systems~\cite{alicki2012, mori2023}. Alternatively, one can look at the Bloch-Redfield equation~\cite{xu2014}, although this equation breaks positivity. While the Markovian property may be broken with time-dependent rates, for sufficiently small oscillations around the constant rates, this property still holds~\cite{maniscalco2014, plenio2010}. 

With this caveat, we now look at how Bloch equations get modified if Eq.~\eqref{eq:time_dependent_equilibrium} is assumed. In the presence of the optical laser and driving, $T$ is assumed to be an effective temperature parameter and not the ambient temperature. 

\subsection{Oscillating equilibrium}

\begin{figure}
    \centering
    \includegraphics[width=.49\textwidth]{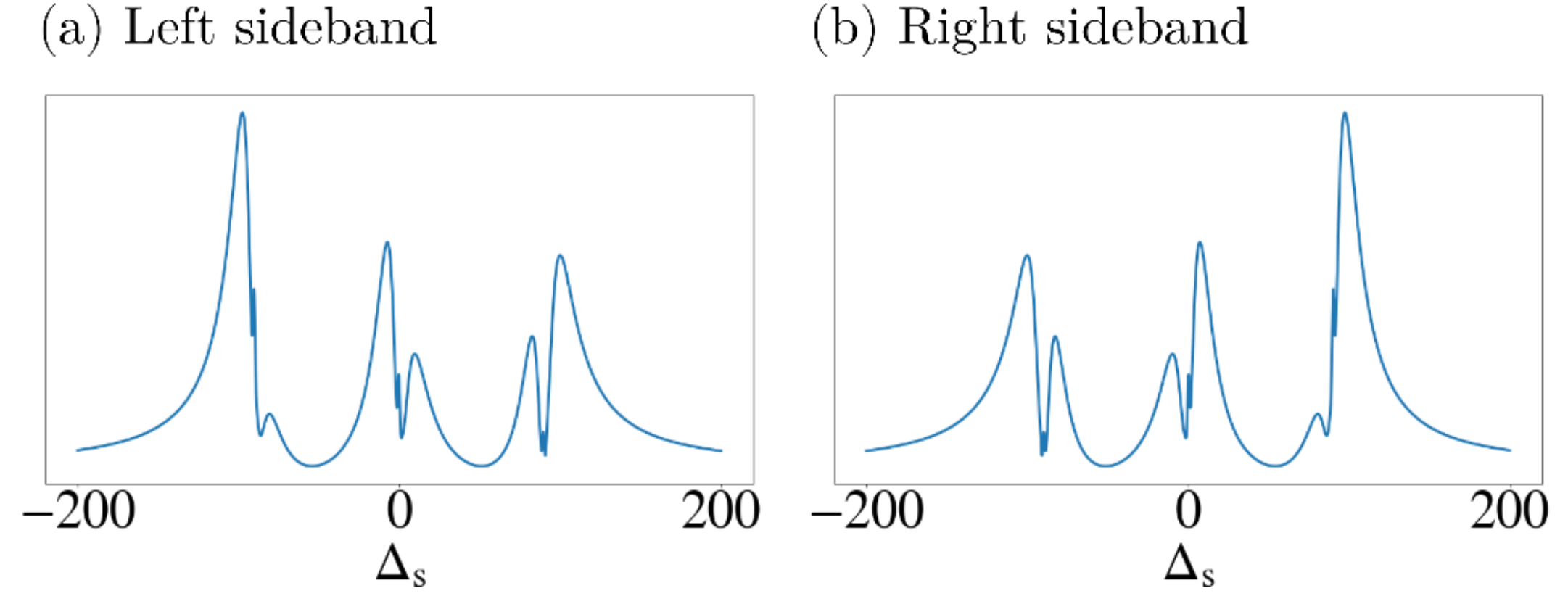}
    \caption{The analytically obtained P1 spectra. The asymmetry of one of the volcanoes is modified because of the interference between the different hyperfine emission lineshapes. The rightmost volcano of the left sideband has the asymmetry opposite to the other two volcanoes, and the same holds for the leftmost volcano of the right sideband. The plots were calculated using Eqs.~\eqref{eq:bath_correction} and \eqref{eq:P1_sigmap}. The parameters used in the plot are $B_{\mathrm{t}}=0.75, \ \omega_{\mathrm{l}}=1.5, \ \gamma_1=0.03, \ \omega_{\mathrm{h}}=90$, and $T=1000$, all in units of $\gamma_2$.}
    \label{fig:p1_spectrum_theory}
\end{figure}

If we assume the time-dependence of the equilibrium polarization as in Eq.~\eqref{eq:time_dependent_equilibrium}, $\sigma_{\mathrm{z}}^{\mathrm{eq}}\left(t\right)$ is to linear order in $B_{\mathrm{l}}$
\begin{equation}
    \sigma_{\mathrm{z}}^{\mathrm{eq}}\left(t\right)=\sigma_{\mathrm{z}, 0}^{\mathrm{eq}}+B_{\mathrm{l}}\left(\sigma_{\mathrm{z}, 1}^{\mathrm{eq}}e^{i\omega_{\mathrm{l}}t}+\sigma_{\mathrm{z}, -1}^{\mathrm{eq}}e^{-i\omega_{\mathrm{l}}t}\right)+\mathcal{O}\left(B_{\mathrm{l}}^2\right).
\end{equation}
If this expression is inserted into the Bloch equations the right sideband solution is modified according to  
\begin{align}\label{eq:bath_correction} &\langle\sigma_+\rangle^{(1)}\rightarrow\langle\sigma_+\rangle^{(1)}+\langle\sigma_+\rangle^{(1)}_{\mathrm{corr}},\\ \nonumber &\langle\sigma_+\rangle^{(1)}_{\mathrm{corr}}=\frac{-i\gamma_1B_t\left(\Delta_{\mathrm{s}}+\omega_{\mathrm{l}}-i\gamma_2\right)\sigma_{\mathrm{z},1}^{\mathrm{eq}}}{(\omega_{\mathrm{l}}-i\gamma_1)(\Delta_{\mathrm{s}}^2-(\omega_{\mathrm{l}}-i\gamma_2)^2)+4|B_\mathrm{t}|^2(\omega_{\mathrm{l}}-i\gamma_2)}.
\end{align}
The P1 right-sideband spectrum with three hyperfine transitions is calculated with the modified expression as
\begin{equation}\label{eq:P1_sigmap}
    \langle\sigma_+\rangle^{(1)}_{\mathrm{P1}}=\sum_{k=\{0,\pm 1\}} \langle\sigma_+\rangle^{(1)}\left(\Delta_{\mathrm{s}}+k\omega_{\mathrm{h}}\right),
\end{equation}
where $\omega_{\mathrm{h}}$ is the hyperfine splitting as seen in Fig.~\ref{fig:p1_hyperfine}. The P1 spectra calculated using this expression are shown in Fig.~\ref{fig:p1_spectrum_theory}. The modified Bloch equations lead to the inversion of asymmetry of one of the volcanoes in the same manner as in the experimentally obtained spectra in Fig.~\ref{fig:experimental_volcano}. 

Lastly, we present the calculation of $\sigma_{\mathrm{z}}^{\mathrm{eq}}(t)$. The exponential of the Hamiltonian can be expanded in the basis of Pauli matrices as 
\begin{equation}
    e^{-\beta|H|\hat{\mathbf{h}}\cdot\boldsymbol{\sigma}}=\cosh\left(\beta|H|\right)\mathbbm{1}-\sinh\left(\beta|H|\right)\hat{\mathbf{h}}\cdot\boldsymbol{\sigma},
\end{equation}
where $|H|$ is the instantaneous eigenvalue of the Hamiltonian, and $\hat{\mathbf{h}}$ is the unit vector in the Pauli basis,
\begin{align}
    |H|(t)&=\frac{1}{2}\sqrt{\left(\Delta_{\mathrm{s}}+B_{\mathrm{l}}\cos\left(\omega_{\mathrm{l}}t\right)\right)^2+4|B_{\mathrm{t}}|^2},\\
    \hat{\mathbf{h}}(t)&=\frac{H_i(t)}{|H|(t)}.
\end{align}
Using the properties of the Pauli matrices, we write Eq.~\eqref{eq:time_dependent_equilibrium} as a function of the previous two parameters,
\begin{equation}
    \sigma_\mathrm{z}^{\mathrm{eq}}(t)=-\tanh{\left(\beta |H|(t)\right)}\hat{\mathbf{h}}_\mathrm{z},
\end{equation}
where $\hat{\mathbf{h}}_\mathrm{z}=\left(\Delta_{\mathrm{s}}+B_{\mathrm{l}}\cos\left(\omega_{\mathrm{l}}t\right)\right)/(2|H|)$. We expand $\sigma_\mathrm{z}^{\mathrm{eq}}(t)$ to linear order in $B_\mathrm{l}$ to obtain
\begin{equation}
    \begin{split}
    &\sigma_{\mathrm{z}}^{\mathrm{eq}}\left(t\right)=-\frac{\Delta_{\mathrm{s}}}{\Omega_\mathrm{R}}\tanh\left(\frac{\beta\Omega_\mathrm{R}}{2}\right)-\frac{B_\mathrm{l}}{\Omega_\mathrm{R}}\cos{\left(\omega_\mathrm{l} t\right)}\\
    &\times\left(\frac{4|B_{\mathrm{t}}|^2}{\Omega_{\mathrm{R}}^2}\tanh\left(\frac{\beta \Omega_\mathrm{R}}{2}\right)+\frac{\beta\Delta_\mathrm{s}^2}{2\Omega_{\mathrm{R}}\cosh^2\left(\frac{\beta\Omega_\mathrm{R}}{2}\right)}\right),
    \end{split}
\end{equation}
where $\Omega_{\mathrm{R}}=\sqrt{\Delta_{\mathrm{s}}^2+4|B_{\mathrm{t}}^2|}$ is the generalized Rabi frequency. In the experiments where $\Omega_{\mathrm{R}}$ is in the MHz range, the thermal energy is much higher than $\Omega_{\mathrm{R}}$, except for very low effective temperatures. The approximate expression in this limit is
\begin{equation}
    \sigma_\mathrm{z}^{\mathrm{eq}}(t)\underset{\frac{\beta}{\Omega_\mathrm{R}}\ll1}{\sim}-\frac{1}{2 k_\mathrm{B} T}\left(\Delta_{\mathrm{s}}+B_\mathrm{l}\cos{(\omega_\mathrm{l} t)}\right).
\end{equation}
 
\bibliography{sideband_spectroscopy}
\end{document}